\documentclass[cmp]{svjour}  
\usepackage{amsmath}
\usepackage{amsfonts,amssymb}

\let\vec\relax
\DeclareMathAccent{\vec}{\mathord}{letters}{"7E}

\newcommand{\bR}{{\mathbb R}}

\newcommand{\kI}{{\mathcal I}}

\newcommand{\kN}{{\mathcal N}}

\newcommand{\gotB}{{\mathfrak B}}
\newcommand{\gotb}{{\mathfrak b}}

\newcommand{\gotf}{{\mathfrak f}}

\newcommand{\gotH}{{\mathfrak H}}
\newcommand{\goth}{{\mathfrak h}}

\newcommand{\ga}{{\alpha}}
\newcommand{\gb}{{\beta}}
\newcommand{\gd}{{\delta}}

\newcommand{\gL}{{\Lambda}}

\newcommand{\gl}{{\lambda}}

\newcommand{\go}{{\omega}}
\newcommand{\gO}{{\Omega}}

\newcommand{\gr}{{\varrho}}

\newcommand{\gs}{{\sigma}}
\newcommand{\gt}{{\tau}}

\newcommand{\slim}{\,\mbox{\rm s-}\hspace{-2pt} \lim}

\newcommand{\dom}{{\mathrm{Dom\,}}}
\newcommand{\ran}{{\mathrm{Ran\,}}}

\newcommand{\tr}{{\mathrm{Tr}}}

\newtheorem{thm}{Theorem}[section]
\newtheorem{prop}[thm]{Proposition}
\newtheorem{lem}[thm]{Lemma}
\newtheorem{cor}[thm]{Corollary}

\newtheorem{defn}[thm]{Definition}
\newtheorem{rem}[thm]{Remark}

\newcommand{\ba}{\begin{array}}
\newcommand{\ea}{\end{array}}
\newcommand{\bea}{\begin{eqnarray}}
\newcommand{\eea}{\end{eqnarray}}
\newcommand{\bead}{\begin{eqnarray*}}
\newcommand{\eead}{\end{eqnarray*}}
\newcommand{\be}{\begin{equation}}
\newcommand{\ee}{\end{equation}}
\newcommand{\bed}{\begin{displaymath}}
\newcommand{\eed}{\end{displaymath}}
\newcommand{\bl}{\begin{lem}}
\newcommand{\el}{\end{lem}}
\newcommand{\bp}{\begin{prop}}
\newcommand{\ep}{\end{prop}}
\newcommand{\bt}{\begin{thm}}
\newcommand{\et}{\end{thm}}
\newcommand{\Label}{\label}
\newcommand{\bc}{\begin{cor}}
\newcommand{\ec}{\end{cor}}
\newcommand{\la}{\Label}

\newcommand{\br}{\begin{rem}}
\newcommand{\er}{\end{rem}}
\newcommand{\bd}{\begin{defn}}
\newcommand{\ed}{\end{defn}}

      \def\dC{{\mathbb C}}
   \def\dE{{\mathbb E}}

   \def\dN{{\mathbb N}}   
      \def\dR{{\mathbb R}}

\numberwithin{equation}{section}

\newcommand{\Imag}{\mbox{{\rm Im}}\,}

\begin{document}

\title{{The effect of time-dependent coupling on non-equilibrium steady states}}

\titlerunning{{Non-equilibrium steady states}}

\author{Horia D. Cornean\inst{1}
\and
Hagen Neidhardt\inst{2}
\and
Valentin Zagrebnov\inst{3}
}
\institute{
Department of Mathematical Sciences, Aalborg University, 
Fredrik Bajers Vej 7G, DK-9220 Aalborg, Denmark\\
\email{cornean@math.aau.dk} \and
WIAS Berlin, Mohrenstr. 39, D-10117 Berlin, Germany\\
\email{neidhard@wias-berlin.de} \and
Universit\'{e} de la M\'{e}diterran\'{e}e 
(Aix-Marseille II) and
Centre de Physique Th\'{e}orique - UMR 6207,
Luminy - Case 907, F-13288 Marseille Cedex 9, France\\
\email{zagrebnov@cpt.univ-mrs.fr}
}

\authorrunning{H.D.~Cornean, H.~Neidhardt, V.~A.~Zagrebnov}

\maketitle

\begin{abstract}
Consider (for simplicity) two one-dimensional semi-infinite leads
cou\-pled to a quantum well via time dependent point interactions. In
the remote past the system is decoupled, and each of its components is at
thermal equilibrium. In the remote future
the system is fully coupled. We define and compute the non equilibrium
steady state (NESS) generated by this evolution. We show that when
restricted to the subspace of absolute continuity of the fully coupled
system, the state does not depend at all on the switching. Moreover, we
show that the stationary charge current has the same invariant 
property, and derive the Landau-Lifschitz and Landauer-B{\"u}ttiker
formulas.
\end{abstract}

\section{Introduction}

The goal of this paper is to construct and study non equilibrium steady
states for systems containing quantum
wells, and to describe the quantum transport of electrons through them. Even
though our results can be generalized to higher dimensions, we choose
for the moment to work in a (quasi) one dimensional setting; let us
describe it in some more detail.

A quantum well
consists of potential barriers which are supposed to confine
particles. On both sides of the barriers are reservoirs of
electrons. Carriers can pass through the barriers by tunneling. 
We are interested in the carrier transport through the barriers, as
well as in the carrier distribution between these barriers. 
Models of such type are very often used to describe processes going on in
nanoelectronic devices: quantum well lasers,
resonant tunneling diodes, and nanotransistors, see \cite{WV1}.

The quasi one dimensional geometry assumes that the carriers can freely move in
the plane orthogonal to the transport axis, but these degrees of
freedom are integrated out. Thus we are
dealing with an essentially one-dimensional physical system. To describe
such a system we consider the transport model of a single band
in a given spatially varying potential $v$, under the assumption
that $v$ and all other possible parameters of the model are
constant outside a fixed interval $(a,b)$, see \cite{Fr1,Fr2,KL1}.

More precisely, in the Hilbert space $\gotH := L^2(\dR)$ we consider the Schr\"odinger operator
\be\la{1.1}
(H f)(x) := -\frac{1}{2}\frac{d}{dx}\frac{1}{M(x)}\frac{d}{dx}f(x) + V(x)f(x),
\quad x \in \dR,
\ee
with domain
\be\la{1.2}
\dom(H) := \{f \in W^{1,2}(\bR): \frac{1}{M}f' \in W^{1,2}(\bR)\}.
\ee
It is assumed that the effective mass $M(x)$ and the real potential
$V(x)$ admit decompositions of the form
\be\la{1.3}
M(x) := 
\begin{cases}
m_a & x \in (-\infty,a]\\
m(x) & x \in (a,b)\\
m_b & x \in [b,\infty)
\end{cases},
\ee
$0 < m_a,m_b < \infty$, $m(x) > 0$, $x \in (a,b)$, $m + \frac{1}{m_{a(b)}}
\in L^\infty((a,b))$,
and
\be\la{1.4}
V(x) := 
\begin{cases}
v_a & x \in (-\infty,a]\\
v(x) & x \in (a,b)\\
v_b & x \in [b,\infty)
\end{cases},\quad v_a \geq  v_b,
\ee
$v_a,v_b \in \bR$, $v \in L^\infty((a,b))$. The quantum well is
identified with the interval $(a,b)$, (or physically, with the
three-dimensional region $(a,b) \times \dR^2$). The regions
$(-\infty,a)$ and $(b,\infty)$ (or physically $(-\infty,a) \times
\dR^2$ and  $(b,\infty) \times \dR^2$), are the reservoirs.

Schr\"odinger operators
with step-like potentials were firstly considered by Buslaev and Fomin in \cite{BF1}.
For that reason we call them Buslaev-Fomin operators. 

The inverse scattering problem for such
Buslaev-Fomin operators was subsequently investigated in 
\cite{AJ1,A1,A2,C1,CK1,GNP1}. 

In order to rigorously describe quantum transport in mesoscopic
systems, these operators were firstly used by P\"otz, see  \cite{P1}. 
In \cite{BKNR1}, the Buslaev-Fomin operator was an important ingredient for
a self-consistent quantum transmitting Schr\"odinger-Poisson system, 
which was used to describe quantum transport in tunneling diodes.
In a further step, this was extended to a so-called hybrid
model which consists of a classical drift-diffusion part and a
quantum transmitting Schr\"odinger-Poisson part, see
\cite{BNR1}. Hybrid models are effective tools of describing and
calculating nanostructures like tunneling diodes, see \cite{BBDE1}.

To obtain a self-consistent description of carrier transport through 
quantum wells, one needs to know the carrier distribution between the barriers
in order to put it into the Poisson equation for determining the
electric field. Semiconductor devices are often modeled in this
manner, see \cite{G1,M1,S1}. Important for that is a relation 
which assigns to each real potential $v \in
L^\infty((a,b))$ a carrier density $u \in L^1((a,b))$. The (nonlinear)
operator doing this is
called the carrier density operator and is denoted by 

$$\kN(\cdot): L^\infty((a,b)) \longrightarrow L^1((a,b)),\qquad \kN(v)
= u.$$

The problem of defining carrier density operators is reduced to the
problem of finding appropriate density operators $\gr$.
\begin{defn}\label{defunu}
A bounded
non-negative operator $\gr$ in $L^2(\bR)$ is called a density operator or a state
if the product $\gr M(\chi_{(a,b)})$ 
is a trace class operator, where $M(\chi_{(a,b)})$ is the multiplication
operator induced in $L^2(\bR)$ by the characteristic function
$\chi_{(a,b)}$ of the interval $(a,b)$.
\end{defn}
 We note that  in general a
non-negative bounded operator is called a state if the operator itself
is a trace class operator and is normalized to one, that is, ${\rm Tr}(\gr)
= 1$. In our case these conditions are relaxed to
the condition that the product $\gr M(\chi_{(a,b)})$ has to be trace
class.  

This weakening is necessary since we are interested in
so-called steady density operators or steady states for Hamiltonians
with continuous spectrum.
\begin{defn}\label{defdoi}
A state $\gr$ is called a {\rm steady state} for $H$ if $\gr$ commutes with $H$,
i.e. $\gr$ belongs to the commutant of the algebra generated by the
functional calculus associated to $H$. A steady state is an {\rm
  equilibrium state} if it belongs to the bicommutant of this algebra.
\end{defn}
Thus if $H$ admits continuous
spectrum, then a steady state cannot be of trace class unless it
equals zero on the subspace of absolute continuity.

To give a description of all possible steady states, one has to introduce
the spectral representation of $H$. Taking into account results of
\cite{BKNR1}, it turns out that the operator $H$ is unitarily
equivalent to the multiplication $M$ induced by the independent
variable $\gl$ in the direct integral $L^2(\bR,\goth(\gl),\nu)$,
\be\label{gothich}
\goth(\gl) := 
\begin{cases}
\dC, & \gl \in (-\infty,v_a]\\
\dC^2, & \gl \in (v_a,\infty)
\end{cases},
\ee
and (with the usual abuse of notation)
\bed 
d\nu(\gl) = \sum_{j=1}^N\gd(\gl - \gl_j)d\gl + \chi_{[v_b,\infty)}(\gl)d\gl, \quad \gl  \in \dR,
\eed
where it is assumed $v_a \ge v_b$, and $\{\gl_j\}^N_{j=1}$ 
denote the finite number of simple eigenvalues of $H$
which are all situated below the threshold $v_b$. We note that
\bed
L^2(\dR,\goth(\gl),\nu) \simeq  \oplus_{j=1}^N\dC \oplus
L^2([v_b,v_a],\dC) \oplus L^2((v_a,\infty),\dC^2).
\eed
The unitary operator
$\Phi: L^2(\dR) \longrightarrow L^2(\dR,\goth(\gl),\nu)$ establishing
the unitary equivalence  of $H$ and $M$ is called the generalized
Fourier transform.

If $\gr$ is a steady state for $H$, then there exists a
$\nu$-measurable function 
$$\dR\ni\gl \mapsto \tilde{\rho}(\gl)\in B(\goth(\gl))$$ of
non-negative bounded operators in $\goth(\gl)$ 
such that $\nu-\sup_{\gl \in \dR}\|\tilde{\rho}(\gl)\|_{\gotB(\goth(\gl))} < \infty$ 
and $\gr$ is unitarily equivalent to the multiplication operator
$M(\tilde{\rho})$ induced by $\tilde{\rho}$ via the generalized Fourier transform
\be\la{1.12}
\gr = \Phi^{-1} M(\tilde{\rho})\Phi.
\ee
The measurable family $\{\tilde{\rho}(\gl)\}_{\gl \in \dR}$ is uniquely
determined by the steady state $\gr$ up to a $\nu$-zero set
and is called the  {\rm distribution function} of the steady state. In other words, there is an
one-to-one correspondence between the set of steady states and the set
of distribution functions. When $\rho$ is an equilibrium state, then
$\tilde{\rho}(\gl)$ must be proportional with the identity operator in $\goth(\gl)$, 
hence $\rho$ must be
a function of $H$. Let us note that the same distribution
function can produce quite different steady states in $L^2(\bR)$. This is due to the fact
that the generalized Fourier transform strongly dependents on $H$, in
particular, on the potential $v$.

Having a steady state $\gr$ for $H$ one defines the carrier density in
accordance with \cite{BKNR1} as the Radon-Nikodym derivative of the
Lebesgue continuous measure $\dE(\go)$
\bed
\dE(\go) := {\rm Tr}(\gr M(\chi_\go))
\eed
where $\go$ is a Borel subset of $(a,b)$. The quantity $\dE(\go)$ can
be regarded as the expectation value that the carriers are contained in
$\go$. Therefore the carrier density $u$ is defined by
\bed
u_\rho(x) := \frac{\dE(dx)}{dx} = \frac{{\rm Tr}(\gr M(\chi_{dx}))}{dx}, \quad x \in (a,b).
\eed
The carrier density operator $\kN_\rho(\cdot): L^\infty((a,b))
\longrightarrow L^1((a,b))$ is now defined as
\be\la{1.15}
\kN_\rho(v) := u_\rho(x)
\ee
where $v \in L^\infty((a,b))$ is the potential of the operator
$H$. The steady state $\gr$ is given by \eqref{1.12}.

Therefore the self-consistent description of the carrier transport through
quantum wells is obtained if there is a way to determine physically
relevant distribution
functions $\tilde{\rho}$. One goal of this paper is to propose a
time-dependent procedure allowing to determine those functions.

\subsection{The strategy}

Let us describe the strategy. We start with a completely decoupled system 
which consists of three subsystems living in the Hilbert spaces
\be\la{1.16}
\gotH_a := L^2((-\infty,a]), \quad
\gotH_\kI := L^2(\kI), \quad
\gotH_b := L^2([b,\infty))
\ee
where $\kI = (a,b)$. We note that
\be\la{1.17}
\gotH = \gotH_a \oplus \gotH_\kI \oplus \gotH_b.
\ee
With $\gotH_a$ we associate the Hamiltonian $H_a$
\bea
(H_a f)(x) & := & -\frac{1}{2m_a}\frac{d^2}{dx^2}f(x) + v_af(x),\la{1.18}\\
f \in \dom(H_a) & := & \{f \in W^{2,2}((-\infty,a)): f(a) = 0\}\la{1.19}
\eea
with $\gotH_\kI$ the Hamiltonian $H_\kI$,
\bea
(H_\kI f)(x) & := &
-\frac{1}{2}\frac{d}{dx}\frac{1}{m(x)}\frac{d}{dx}f(x) + v(x)f(x),\la{1.20}\\
f \in \dom(H_\kI) & := & \left\{f \in W^{1,2}(\kI): 
\ba{l}
\frac{1}{m}f' \in W^{1,2}(\kI)\\
f(a) = f(b) = 0
\ea
\right\}\la{1.21}
\eea
and with $\gotH_b$ the Hamiltonian $H_b$,
\bea
(H_b f)(x) & := & -\frac{1}{2m_b}\frac{d^2}{dx^2}f(x) + v_b f(x),\la{1.22}\\
f \in \dom(H_b) & := & \{f \in W^{2,2}((b,\infty): f(b) = 0\}.\la{1.23}
\eea
In $\gotH$ we set
\be\la{1.24}
H_D := H_a \oplus H_\kI \oplus H_b
\ee
where the sub-index $``D''$ indicates Dirichlet boundary conditions. The
quantum subsystems $\{\gotH_a,H_a\}$ and $\{\gotH_b,H_b\}$ are called
left- and right-hand reservoirs. The middle system
$\{\gotH_\kI,H_\kI\}$ is identified with a closed quantum well. We
assume that all three subsystems are at thermal equilibrium; according
to Definition \ref{defdoi}, the corresponding sub-states must be functions
of their corresponding sub-Hamiltonians. The total state is the direct
sum of the three sub-states.

One example borrowed from the
physical literature, which takes into account the quasi one dimensional
features of our problem is as follows. Assume the same temperature
$T$. The equilibrium sub-states are $\gr_a$, $\gr_\kI$ and
$\gr_b$ where:
\be\la{1.25}
\gr_a := \gotf_a(H_a - \mu_a), \quad
\gr_\kI := \gotf_\kI(H_\kI - \mu_\kI), \quad 
\gr_b := \gotf_b(H_b - \mu_b)
\ee
where
\bed
\gotf_a(\gl) := c_a\ln(1 + e^{-\gb \gl}), \;
\gotf_\kI(\gl) := c_\kI\ln(1 + e^{-\gb \gl}),\;
\gotf_b(\gl) := c_b\ln(1 + e^{-\gb \gl}),
\eed
$\gl \in \bR$, $\gb := 1/(kT)$, $k$ is the Boltzmann constant, 
$\mu_a$ and $\mu_b$ are the chemical potentials of left- and
right-hand reservoirs and $\mu_\kI$ the chemical potential of the
quantum well. The constants $c_a$, $c_\kI$ and $c_b$ are given by
\be\la{1.26}
c_a := \frac{q\, m^*_a}{\pi\,\gb}, \quad 
c_\kI := \frac{q\, m^*_\kI}{\pi\, \gb}, \quad
c_b := \frac{q\, m^*_b}{\pi\, \gb}
\ee
where $m^*_a$, $m^*_\kI$ and $m^*_b$ are the electronic effective masses
appearing after integrating out the orthogonal degrees of freedom (see
for more details \cite{Fr1,Fr2}).
 
We set
\be\la{1.27}
\gr_D := \gr_a \oplus \gr_\kI \oplus \gr_b.
\ee
For the whole system $\{\gotH,H_D\}$ the state $\gr_D$ is a 
steady state because $\gr_D$ commutes with $H_D$ (see Definition \ref{defdoi}).
In general, the state $\gr_D$ cannot be represented as a function of
$H_D$ which is characteristic for equilibrium states, but it is the
direct sum of equilibrium sub-states. In any case, $\gr_D$ is
a special non-equilibrium steady state (NESS) for the system
$\{\gotH,H_D\}$. Now here comes the main question: can we construct a
NESS for $\{\gotH,H\}$ starting from $\gr_D$?

Let us assume that at $t = -\infty$ the quantum system $\{\gotH,H_D\}$
is described by the NESS $\gr_D$. Then we connect in a time dependent
manner the left- and right-hand reservoirs to
the closed quantum well $\{\gotH_\kI,H_\kI\}$. We assume that the connection process is
described by the time-dependent Hamiltonian
\be\label{hamfamily}
H_\ga(t) := H + e^{-\ga t}\gd(x-a) + e^{-\ga t}\gd(x-b), \quad t \in
\bR, \quad \ga > 0.
\ee
The operator $H_\ga(t)$ is defined by 
\be\la{1.29}
(H_\ga(t))f)(x) := 
-\frac{1}{2}\frac{d}{dx}\frac{1}{M(x)}\frac{d}{dx}f(x) + V(x)f(x), \quad f \in \dom(H_\ga(t)),
\ee
where the domain $\dom(H_\ga(t))$ is given by
\bea\la{1.30}
\lefteqn{
\dom(H_\ga(t)) := }\\
& &
\left\{f \in W^{1,2}(\dR): 
\ba{l}
\frac{1}{M}f' \in W^{1,2}(\dR)\\
(\frac{1}{2M}f')(a+0) - (\frac{1}{2M}f')(a-0) = e^{-\ga t}f(a)\\
(\frac{1}{2M}f')(b+0) - (\frac{1}{2M}f')(b-0) = e^{-\ga t}f(b)
\ea
\right\}.
\nonumber
\eea
After a rather standard analysis, one can prove the following
convergence in the norm resolvent sense:
\be\la{1.31}
n-\lim_{t \to -\infty} (H_\ga(t) - z)^{-1} = (H_D - z)^{-1}
\ee
and
\be\la{1.32}
n-\lim_{t \to +\infty} (H_\ga(t) - z)^{-1} = (H - z)^{-1},
\ee
$z \in \dC \setminus \dR$. Then we consider the quantum
Liouville equation (details about the various topologies will follow later):
\be\la{1.33}
i\frac{\partial}{\partial t}\gr_\ga(t) = [H_\ga(t),\gr_\ga(t)], \quad
t \in \bR,
\ee
for a fixed $\ga > 0$ satisfying the initial condition
\bed
\slim_{t\to-\infty} \gr_\ga(t) = \gr_D.
\eed
Having found a solution $\gr_\ga(t)$ we are interested in the ergodic limit
\bed
\gr_\ga = \lim_{T\to+\infty}\frac{1}{T}\int_0^T\gr_\ga(t)dt.
\eed
If we can verify that the limit $\gr_\ga$ exists and commutes with
$H$, then $\gr_\ga$ is regarded as the desired NESS of the fully
coupled system $\{\gotH,H\}$. Inserting $\gr_\ga$
into the definition of the carrier density operator $\kN_{\rho_\ga}$ 
we complete the definition of the carrier density operator. Finally,
the steady state $\gr_a$ allows to determine the corresponding
distribution function $\{\tilde{\rho}_\ga(\gl)\}_{\gl \in \dR}$.

\subsection{Outline of results}

The precise formulation of our main result can be found in Theorem
\ref{maintheorem}, and here we only describe its main
features in words. 

We need to introduce the incoming wave operator
\be\label{incwaveop}
W_- := \slim_{t\to-\infty}e^{itH}e^{-itH_D}P^{ac}(H_D)
\ee
where $P^{ac}(H_D)$ is the projection on the absolutely
continuous subspace $\gotH^{ac}(H_D)$ of $H_D$. We note 
that $\gotH^{ac}(H_D) = L^2((-\infty,a]) \oplus L^2([b,\infty))$.
The wave operator exists and is complete, that is, $W_-$ is an
isometric operator acting from $\gotH^{ac}(H_D)$ onto $\gotH^{ac}(H)$
where $\gotH^{ac}(H)$ is the absolutely continuous subspace
of $H$ (the range of $P^{ac}(H)$).

One not so surprising result, is that $\gr_\ga$ exists for all
$\alpha>0$. In fact, if we restrict ourselves to the subspace
$\gotH^{ac}(H)$, then we do not need to take the ergodic limit, since
the usual strong limit exist. 
The surprising fact is that
\begin{equation}\label{surprise}
\slim_{t\to\infty}\gr_\ga (t)P^{ac}(H)=\gr_\ga P^{ac}(H)=W_-\rho_DW_-^*P^{ac}(H),
\end{equation}
which is {\it independent} of $\alpha$. 

The only $\alpha$ dependence
can be found in $\gr_\ga P^{d}(H)$, where $P^{d}(H)$ is the projection
on the subspace generated by the discrete eigenfunctions of $H$. But
this part does not contribute to the stationary current as can be seen
in Section \ref{section4}. Here the ergodic limit is essential, because it kills off
the oscillations produced by the interference between different
eigenfunctions.

Note that the case $\alpha=\infty$ would describe the
situation in which the coupling is suddenly made at $t=0$ and then the
system evolves freely with the dynamics generated by $H$ (see
\cite{JP}, and the end of Section \ref{section3}). 

The case $\alpha\searrow 0$ would correspond to the adiabatic
limit. Inspired by the physical literature which seems to claim that
the adiabatic limit would take care of the above mentioned
oscillations, we {\it conjecture} the following result for the
transient current:
\begin{conjecture}\label{conjectura}
$$\lim_{\alpha\searrow 0}\limsup_{t\to \infty}\left \vert {\rm Tr} \{\gr_\ga
(t)P^{d}(H)[H,\chi]\}\right \vert =0,$$
where $\chi$ is any smoothed out characteristic function of one of the reservoirs.
\end{conjecture}

Before ending this introduction, let us comment on some other physical aspects related 
to quantum transport problems. Many physics papers are dealing with transient currents and not 
only with the steady ones. More precisely, they investigate non-stationary electronic transport 
in noninteracting nanostructures driven by a finite bias and time-dependent signals 
applied at their contacts to the leads, while they allow the carriers to self-interact inside the 
quantum well (see for example \cite{MGM}, \cite{NST} and references therein). 

An interesting open problem is to study the existence of NESS in the Cini (partition-free) 
approach \cite{Cini}, \cite{CJM1}, \cite{CJM2}, \cite{CJM3}. 
Some nice results which are in the same spirit with ours 
have already been obtained in the physical literature \cite{Ste}, \cite{SA}, even 
for systems which 
allow local self-interactions.

\vspace{0.5cm}

Now let us describe the organization of our paper. 

Section \ref{section2} introduces all the necessary notation and
presents an explicit description of a
spectral representation of $H_D$ and $H$. 

Section \ref{section3} deals with the quantum Liouville equation,
and contains the proof of our main result in Theorem \ref{maintheorem}.  

In Section \ref{section4} we define the stationary current and derive the
Landau-Lifschitz and Landauer-B\"uttiker formulas. 

\section{Technical preliminaries}\label{section2}

\subsection{The uncoupled system}

Let us start by describing the uncoupled system, and begin with the
left reservoir. The spectrum of $H_a$ is absolutely continuous
and $\gs(H_a) = \gs_{ac}(H_a) = [v_a,\infty)$. The operator is
simple. The generalized eigenfunctions $\psi_a(\cdot,\gl)$, $\gl \in
[v_a,\infty)$, of $H_a$ are given by
\bed
\psi_a(x,\gl) := \frac{\sin(2m_a q_a(\gl)(x-a)))}{\sqrt{\pi q_a(\gl)}},
\quad x \in (-\infty,a], \quad \gl \in [v_a,\infty)
\eed
where
\bed
q_a(\gl) := \sqrt{\frac{\gl - v_a}{2m_a}}.
\eed
The system of eigenfunctions $\{\psi_a(\cdot,\gl)\}_{\gl \in [v_a,\infty)}$
is orthonormal, that is, one has in distributional sense 
\be\la{2.3}
\int^\infty_a dx \;\psi_a(x,\gl)\overline{\psi_a(x,\mu)} = \gd(\gl - \mu), 
\quad \gl,\mu \in [v_a,\infty).
\ee
With the generalized eigenfunctions one associates the generalized
Fourier transform $\Psi_a: L^2((-\infty,a]) \longrightarrow
L^2([v_a,\infty))$ given by 
\bed
(\Psi_a f)(\gl) = \int^a_{-\infty} f(x) \overline{\psi_a(x,\gl)} dx =
\int^a_{-\infty} f(x) \psi_a(x,\gl) dx, 
\eed 
$f \in L^2((-\infty,a])$. Using \eqref{2.3} a straightforward computation shows that the
generalized Fourier is an isometry acting from $L^2((-\infty,a])$ onto
$L^2([v_a,\infty))$. The inverse operator 
$\Psi^{-1}_a: L^2([v_a,\infty) \longrightarrow L^2(-\infty,a])$ admits
the representation
\bed
(\Psi^{-1}_af)(\gl) = \int^\infty_{v_a} \psi_a(x,\gl)f(\gl)d\gl
=
\int^\infty_{v_a}\frac{\sin(2m_a q_a(\gl)(x-a)))}{\sqrt{\pi q_a(\gl)}}f(\gl)d\gl,
\eed
$f \in L^2([v_a,\infty))$. Since $\psi_a(\cdot,\gl)$ are generalized
eigenfunctions of $H_a$ one easily verifies that
\bed
M_a = \Psi_a H_a \Psi^{-1}_a 
\quad \mbox{or} \quad
H_a = \Psi^{-1}_a M_a \Psi_a
\eed
where $M_a$ is the multiplication operator induced by the independent
variable $\gl$ in $L^2([v_a,\infty))$ and defined by
\bead
(M_af)(\gl) & = & \gl f(\gl), \\
f \in \dom(M_a) & := & \{f \in
L^2([v_a,\infty)): \gl f(\gl) \in L^2([v_a,\infty))\}.
\eead
This shows that $\{L^2([v_a,\infty)),M_a\}$ is a spectral
representation of $H_a$. For the equilibrium sub-state $\gr_a = \gotf_a(H_a -
\mu_a)$ one has the representation
\bed
\gr_a = \Psi^{-1}_a M(\gotf_a(\cdot - \mu_a))\Psi_a
\eed
where $M(\gotf_a(\cdot - \mu_a))$ is the multiplication operator induced by the function
$\gotf_a(\cdot - \mu_a)$.

\vspace{0.5cm}

Let us continue with the closed quantum well. The operator $H_\kI$ has purely discrete point spectrum $\{\xi_k\}_{k \in \dN}$
with an accumulation point at $+\infty$. The eigenvalues are simple.
The density operator $\gr_\kI = \gotf_\kI(H_\kI - \mu_\kI)$ is trace
class. One easily verifies that there is an isometric map
$\Psi_\kI : L^2((a,b)) \longrightarrow L^2(\dR,\dC,\nu_\kI)$,
$d\nu_\kI(\gl) = \sum^\infty_{k=1}\gd(\gl - \xi_k)d\gl$, such that
$\{L^2(\dR,\dC,\nu_\kI),M_\kI\}$ becomes a spectral 
representation of $H_\kI$ where $M_\kI$ denotes the multiplication operator
in $L^2(\dR,\dC,\nu_\kI)$.

\vspace{0.5cm}

Finally, the right-hand reservoir. The spectrum of $H_b$ is
absolutely continuous and $\gs(H_b) = \gs_{ac}(H_b) = [v_b,\infty)$. The
operator $H_b$ is simple. The generalized eigenfunctions
$\psi_b(\cdot,\gl)$, $\gl \in [v_b,\infty)$ are given by
\bed
\psi_b(x,\gl) = \frac{\sin(2m_bq_b(\gl)(x-b))}{\sqrt{\pi q_b(\gl)}},
\quad x \in [b,\infty), \quad \gl \in [v_b,\infty)
\eed
where
\bed
q_b(\gl) = \sqrt{\frac{\gl - v_b}{2m_b}}.
\eed
The generalized eigenfunctions 
$\{\psi_b(\cdot,\gl)\}_{\gl \in [v_b,\infty)}$ perform an orthonormal system and define 
a generalized Fourier transform $\Psi_b: L^2([b,\infty))
\longrightarrow L^2([v_b,\infty))$ by
\bed
(\Psi_b f)(\gl) := \int^\infty_b f(x)\overline{\psi_b(x,\gl)} dx 
= \int^\infty_b f(x)\psi_b(x,\gl) dx,
\eed
$f \in L^2([b,\infty))$. The inverse Fourier transform $\Psi^{-1}_b:
L^2([v_b,\infty)) \longrightarrow L^2([b,\infty))$ admits the
representation
\bed
(\Psi^{-1}_b f)(x) = \int^\infty_{v_b} \psi_b(x,\gl)f(\gl)d\gl = 
\int^\infty_{v_b} \frac{\sin(2m_bq_b(\gl)(x-b))}{\sqrt{\pi q_b(\gl)}}f(\gl)d\gl,
\eed
$f \in L^2([v_b,\infty))$. Denoting by $M_b$ the multiplication
operator induced by the independent variable $\gl$ in
$L^2([v_b,\infty))$ we get
\bed
M_b = \Psi_b H_b \Psi^{-1}_b 
\quad \mbox{or} \quad
H_b = \Psi^{-1}_b M_b \Psi_b
\eed
which shows that $\{L^2([v_b,\infty)),M_b\}$ is a spectral
representation of $H_b$. The equilibrium sub-state $\gr_b = \gotf_b(H_b - \mu_b)$
is unitarily equivalent to the multiplication operator
$M(\gotf_b(\cdot-\mu_b))$ induced by the function $F(\cdot - \mu_b)$ in
$L^2([v_b,\infty))$, that is,
\bed
\gr_b = \Psi^{-1}_b M(\gotf_b(\cdot - \mu_b))\Psi_b.
\eed

\subsection{Spectral representation of the decoupled system}

A straightforward computation shows that the direct sum $\Psi = \Psi_a \oplus \Psi_\kI
\oplus \Psi_b$ defines an isometric map acting from $L^2(\dR)$ onto 
$L^2(\dR,\goth(\gl),\nu_D(\gl))$, $d\nu_D(\gl) =
\sum^\infty_{k=1}\gd(\gl - \xi_k)d\gl + \chi_{[v_b,\infty)}(\gl)d\gl$,
such that $H_D$ becomes unitarily equivalent to the multiplication
operator $M_D$ defined in $L^2(\dR,\goth(\gl),\nu_D(\gl))$ (see
\eqref{gothich}). Here we slightly change the definition of
$\goth(\gl)$ such that it re-becomes $\mathbb{C}$ when $\lambda$ hits an
eigenvalue. This does not affect the absolutely continuous part. 

Hence $\{L^2(\dR,\goth(\gl),\nu_D(\gl)),M_D\}$ is a spectral
representation of $H_D$. Under the map $\Psi$ the absolutely
continuous part $H^{ac}_D = H_a \oplus H_b$  of $H_D$ is unitarily
equivalent to the multiplication operator $M$ in
$L^2(\bR,\goth(\gl),\nu^{ac}_D\}$, $d\nu^{ac}_D(\gl) =
\chi_{[v_b,\infty)}(\gl)d\gl$. Therefore
$\{L^2(\dR,\goth(\gl),\nu^{ac}_D),M\}$
is a spectral representation of $H^{ac}_D$.

With respect to the spectral representation
$\{L^2(\dR,\goth(\gl),\nu_D),M\}$
the distribution function $\{\tilde{\rho}_D(\gl)\}$ of the steady state $\gr_D$ is given by
\bed
\tilde{\rho}_D(\gl) :=
\begin{cases} 
0, & \gl \in \dR \setminus \gs(H_D)\\
\gotf_\kI(\gl - \mu_\kI), & \gl \in \gs_p(H_D) = \gs(H_\kI)\\
\gotf_b(\gl -\mu_b), & \gl \in [v_b,v_a) \setminus \gs(H_\kI)\\
\begin{pmatrix}
\gotf_b(\gl - \mu_b) & 0 \\
0 & \gotf_a(\gl - \mu_a)
\end{pmatrix}, & \gl \in [v_a,\infty) \setminus \gs(H_\kI)
\end{cases}
\eed
We note that $M(\tilde{\rho}_D) = \Psi \gr_D \Psi^{-1}$.

\subsection{The fully coupled system}

The Hamiltonian $H$ in \eqref{1.1} was investigated in detail in \cite{BKNR1}. If
$v_a \ge v_b$, then it turns out 
that the operator $H$ has a finite simple point spectrum on $(-\infty,v_b)$,
on $[v_b,v_a)$ the spectrum is absolutely continuous and simple, and 
on $[v_a,\infty)$ the spectrum is also absolutely continuous with 
multiplicity two. 

Denoting by $\{\gl_p\}^N_{p=1}$ the eigenvalues on 
$(-\infty,v_b)$, we have a corresponding finite sequence of $L^2$-eigenfunctions
$\{\psi(x,\gl_j)\}^N_{j=1}$. 

Moreover, one can construct a set of generalized
eigenfunctions $\phi_a(x,\gl)$, $x \in \dR$, $\gl \in [v_a,\infty)$,
and $\phi_b(x,\gl)$, $x \in \dR$, $\gl \in [v_b,\infty)$ of $H$ such
that $\{\phi_b(\cdot,\gl)\}_{\gl \in [v_b,v_a)}$ and 
$\{\phi_b(\cdot,\gl),\phi_a(\cdot,\gl)\}_{\gl \in [v_a,\infty)}$
generate a complete orthonormal systems of generalized
eigenfunctions. More precisely:
\bead
\int_\dR \phi_a(x,\gl)\overline{\phi_a(x,\mu)}dx & = &\gd(\gl - \mu), \quad
  \gl,\mu \in [v_a,\infty)\\
\int_\dR \phi_b(x,\gl)\overline{\phi_b(x,\mu)} dx & = & \gd(\gl -
\mu), \quad \gl,\mu \in [v_b,\infty)\\
\int_\dR \phi_a(x,\gl)\overline{\phi_b(x,\mu)} dx & = & 0, \quad
\gl,\mu \in [v_a,\infty),
\eead
see \cite{BKNR1}. The existence of generalized eigenfunctions is shown
by constructing solutions 
$\widetilde{\phi}_a(x,\gl)$ and $\widetilde{\phi}_b(x,\gl)$ of the
ordinary differential equation
\bed
-\frac{1}{2}\frac{d}{dx}\frac{1}{m(x)}\frac{d}{dx}\widetilde{\phi}_p(x,\gl)
+ v(x)\widetilde{\phi}_p(x,\gl) = \gl \widetilde{\phi}_p(x,\gl),
\eed
$x \in \dR$, $\gl \in [v_b,\infty)$, $p = a,b$, obeying
\bed
\widetilde{\phi}_a(x,\gl) =
\begin{cases}
e^{i2m_aq_a(\gl)(x-a)} + S_{aa}(\gl)e^{-i2m_aq_a(\gl)(x-a)} & x \in (-\infty,a]\\
S_{ba}(\gl)e^{i2m_bq_b(\gl)(x-b)} & x \in [b,\infty),
\end{cases}
\eed
$\gl \in [v_a,\infty)$, and
\bed
\widetilde{\phi}_b(x,\gl) =
\begin{cases}
S_{ab}(\gl)e^{-i2m_aq_a(\gl)(x-a)} & x \in (-\infty,a]\\
e^{-i2m_bq_b(\gl)(x-b)} + S_{bb}(\gl)e^{i2m_bq_b(\gl)(x-b)} & x \in [b,\infty),
\end{cases}
\eed
$\gl \in [v_b,\infty)$. The coefficients $S_{aa}(\gl)$ and
$S_{bb}(\gl)$ are called reflection coefficients while $S_{ba}(\gl)$
and $S_{ab}(\gl)$ are called transmission coefficients. The solutions
$\widetilde{\phi}_a(\gl)$ and $\widetilde{\phi}_b(\gl)$ define the
normalized generalized eigenfunctions of $H$ by
\bead
\phi_b(x,\gl) := \frac{1}{4\pi q_b(\gl)}\widetilde{\phi}_b(x,\gl), & x
  \in \dR, \quad \gl \in [v_b,\infty),\\
\phi_a(x,\gl) := \frac{1}{4\pi q_a(\gl)}\widetilde{\phi}_a(x,\gl), & x
  \in \dR, \quad \gl \in [v_a,\infty).
\eead
Having the existence of the generalized eigenfunctions one introduces 
the generalized Fourier transform $\Phi: L^2(\dR)
\longrightarrow L^2(\goth(\dR,\goth(\gl),\nu)$ by
\begin{equation}\la{2.23}
(\Phi f)(\gl) := \int_\dR f(x)\overline{\vec{\phi}(x,\gl)}dx, \quad
f \in L^2(\dR), \quad \gl \in \gs(H),
\end{equation}
where
\be\la{2.24}
\vec{\phi}(x,\gl) :=
\begin{cases}
\phi(x,\gl_j) & \gl \in \gs_p(H), \quad x \in \dR\\
\phi_b(x,\gl) & \gl \in [v_b,v_a), \quad x \in \dR\\
\begin{pmatrix}
\phi_b(x,\gl)\\
\phi_a(x,\gl)
\end{pmatrix} & \gl \in [v_b,\infty),\quad x \in \dR,
\end{cases}
\ee
see \cite{BKNR1}. The inverse generalized Fourier transform
$\Phi^{-1}: L^2(\dR,\goth(\gl),\nu) \longrightarrow L^2(\bR)$ is given
by
\be\la{2.25}
(\Phi^{-1}g) = \int_\dR \langle
g(\gl),\overline{\vec{\phi}(x,\gl)}\rangle _{\goth(\gl)} d\nu(\gl),
\ee
$x \in \dR$, $g \in L^2(\bR,\goth(\gl),\nu)$ where
$\langle \cdot,\cdot\rangle _{\goth(\gl)}$ is the scalar product in $\goth(\gl)$.
Since $\Phi$ is an isometry action from $L^2(\dR)$ onto
$L^2(\dR,\goth(\gl),\nu)$ and
\bed
M = \Phi H \Phi^{-1}
\eed
holds where $M$ is multiplication operator induced by the independent
variable in $L^2(\dR,\goth(\gl),\nu)$ one gets that
$\{L^2(\dR,\goth(\gl),\nu),M\}$ is a spectral representation of
$H$.

Under $\Phi$ the absolutely continuous part $H^{ac}$ becomes unitarily
equivalent to the multiplication operator $M$ in
$L^2(\dR,\goth(\gl),\nu^{ac})$, $d\nu^{ac} =
\chi_{[v_b,\infty)}d\gl$. Hence $\{L^2(\dR,\goth(\gl),\nu^{ac}),M\}$
is a spectral representation of $H^{ac}$, as it was for $H^{ac}_D$.

\subsection{The incoming wave operator}

We have already mentioned that $W_-$ as defined in \eqref{incwaveop}
exists and is complete \cite{Yafaev}. We will need in Section
\ref{section4} the expression of the "rotated" wave operator $\Phi
W_-\Psi^{-1}$ which acts from $L^2(\dR,\goth(\gl),\nu^{ac})$ onto 
itself. By direct (but tedious) computations one can show that 
$\widehat{W}_- := \Phi W_-\Psi^{-1}$ acts as a multiplication
operator, which means that there is a
family $\{\widetilde{W}_-(\gl)\}_{\gl \in \bR}$ of isometries acting from
$\goth(\gl)$ onto $\goth(\gl)$  such that
\bed
(\widehat{W}_-f)(\gl) = \widetilde{W}_-(\gl)f(\gl), \quad f \in
L^2(\bR,\goth(\gl),\nu^{ac}).
\eed
The family $\{\widetilde{W}_-(\gl)\}_{\gl \in \bR}$ is called the incoming wave matrix and
can be explicitly calculated. One gets
\be\la{2.31}
\widetilde{W}_-(\gl) = 
\begin{cases}
i & \gl \in [v_b,v_a]\\
\begin{pmatrix}
i & 0\\
0 & -i
\end{pmatrix} & \gl \in (v_a,\infty).
\end{cases}
\ee

Note that another possible approach to the spectral problem (and
completely different) would be to
construct generalized eigenfunctions for $H$ out of those of $H_D$ by
using the unitarity of $W_-$ between their subspaces of absolute
continuity and the formal intertwining identity
$\phi_p(\cdot,\lambda):=W_-\psi_p(\cdot,\lambda)$. In this case
$\widetilde{W}_-(\gl)$ would always equal the identity matrix. 

\section{The quantum Liouville equation}\label{section3}

The time dependent operators $H_\ga(t)$ from \eqref{hamfamily} are defined by the sesquilinear
forms $\goth_\ga[t](\cdot,\cdot)$,
\bea\label{formfamily}
\lefteqn{
\goth_\ga[t](f,g) = }\\
& &
\int_\dR \left\{f'(x)\overline{g'(x)} + 
  V(x)f(x)\overline{g(x)}\right\} dx + 
e^{-\ga t}f(a)\overline{g(a)} + e^{-\ga t}f(b)\overline{g(b)},
\nonumber
\eea
$f,g \in \dom(\goth_a[t]) := W^{1,2}(\dR)$, $t \in \dR$. Obviously, we
have $H_\ga(t) + \gt \ge I$, $\gt := \|V\|_{L^\infty(\dR)} +
1$. For each $t \in \dR$ the operator $H_\ga(t)$ can be regarded as a
bounded operator acting from $W^{1,2}(\dR)$ into
$W^{-1,2}(\dR)$. Classical Sobolev embedding results ensure that
$(H_\ga(t) + \gt)^{-1/2}$ maps $L^2(\dR)$ into continuous functions,
and it has an integral kernel $G(x,x';\tau)$ with the property that
$G(\cdot ,x';\tau)\in L^2(\dR)$ for every fixed $x'$.
 
Let us introduce the operators $B_a: L^2(\dR) \longrightarrow \dC$
and $B_b: L^2(\dR) \longrightarrow \dC$ defined by:
\bea\label{rangunu}
\lefteqn{
(B_af) := [(H_\ga(t) + \gt)^{-1/2}f](a)=}\\
& &
\int_{\dR}G(a,x;\tau)f(x)dx,\qquad (B_a^*c)(x) =  G(x,a;\tau),
\nonumber
\eea
and similarly for $B_b$. The operators $B_a^*B_a$ and $B_b^*B_b$ are
bounded in $L^2(\dR)$ and correspond to the sesquilinear forms
\bed
\gotb_a[t](f,g) := ((H_\ga(t) + \gt)^{-1/2}f)(a)\overline{((H_\ga(t) + \gt)^{-1/2}g)(a)},
\eed
$f,g \in \dom(\gotb_a[t]) = L^2(\dR)$, and 
\bed
\gotb_b[t](f,g) := ((H_\ga(t) + \gt)^{-1/2}f)(b)\overline{((H_\ga(t) + \gt)^{-1/2}g)(b)},
\eed
$f,g \in \dom(\gotb_b[t]) = L^2(\dR)$, respectively. We define the
rank two operator
\bea\label{bebebe}
\lefteqn{
B := B^*_aB_a + B^*_bB_b=}\\
& &
G(\cdot ,a;\tau)G(a,\cdot;\tau)+G(\cdot ,b;\tau)G(b,\cdot;\tau): L^2(\dR) \longrightarrow L^2(\dR).
\nonumber
\eea
The resolvent $(H_\ga(t) + \gt)^{-1}$ admits the representation
\be\la{2.35}
(H_\ga(t) + \gt)^{-1} = (H + \gt)^{-1/2}(I + e^{-\ga t}B)^{-1}(H +
\gt)^{-1/2}, \quad t \in \dR, \quad \ga > 0.
\ee

\subsection{The unitary evolution}

Let us consider a weakly differentiable map $\mathbb{R}\ni t\mapsto u(t)\in
W^{1,2}(\dR)$. We are interested in the evolution equation
\be\la{2.36}
i\frac{\partial}{\partial t}u(t) = H_\ga(t)u(t), \quad t \in \dR,
\quad \ga > 0.
\ee
where $H_\ga(t)$ is regarded as a bounded operator acting from $W^{1,2}(\dR)$
into $W^{-1,2}(\dR)$.

By Theorem 6.1 of \cite{NZ1} with evolution equation \eqref{2.36} one
can associate a unique
unitary solution operator or propagator $\{U(t,s)\}_{(t,s) \in \dR \times \dR}$ leaving
invariant the Hilbert space $W^{1,2}(\dR)$. By Theorem 8.1 of
\cite{Ki1} we find that for $x,y \in W^{1,2}(\dR)$ the sesquilinear
form $(U(t,s)x,y)$ is continuously differentiable with respect $t \in
\dR$ and $s \in \dR$ such that
\bea
\frac{\partial}{\partial t}(U(t,s)x,y) & = & -i(H_\ga(t)U(t,s)x,y), \quad
x,y \in W^{1,2}(\dR),\la{2.37}\\
\frac{\partial}{\partial s}(U(t,s)x,y) & = & i(H_\ga(s)x,U(s,t)y), \quad
x,y \in W^{1,2}(\dR).\la{2.38}
\eea

\subsection{Quantum Liouville equation}

We note that 
\be\la{3.9}
\gr_\ga(t) := U(t,s)\gr_\ga(s) U(s,t), \quad t,s \in \dR,
\ee
seen as a map from $W^{1,2}(\dR)$ into $W^{-1,2}(\dR)$ is
differentiable and solves the quantum Liouville equation \eqref{1.33}
satisfying the initial condition $\gr_\ga(t)|_{t = s} = \gr_\ga(s)$,  
provided $\gr_\ga(s)$ leaves 
$W^{1,2}(\dR)$ invariant. Indeed, using \eqref{2.37} and \eqref{2.38} we find
\bead
\lefteqn{
\frac{\partial}{\partial t}(\gr_\ga(s)U(s,t)x,U(s,t)y) =}\\
& & 
i(U(s,t)H_\ga(t)x,\gr_\ga(s)U(s,t)y) - i((\gr_\ga(s)U(s,t)x,U(s,t)H_\ga(t)y) = 
\nonumber\\
& &
i(H_\ga(t)x,\gr_\ga(t)y) - i(\gr_\ga(t)x,H_\ga(t)y),
\eead
$x,y \in W^{1,2}(\dR)$, which yields
\bed
i\frac{\partial}{\partial t}(\gr_\ga(t)x,y)
= (\gr_\ga(t)x,H_\ga(t)y) - (H_\ga(t)x,\gr_\ga(t)y),
\eed
$x,y \in W^{1,2}(\dR)$, $t,s \in \dR$.

\subsection{Time dependent scattering}

We set $U(t) := U(t,0)$, $t \in \dR$  and consider the wave operators
\bed
\gO_- := \slim_{t\to-\infty}U(t)^*e^{-itH_D}
\eed
and
\bed
\gO_+ := \slim_{t\to+\infty}U(t)^*e^{-itH}.
\eed
\bp
Let $H_D$ and $H_\ga(t)$, $t \in \dR$, $\ga > 0$, be given by 
\eqref{1.16}-\eqref{1.24} and \eqref{1.29}-\eqref{1.30}, respectively.
Then the wave operator $\gO_-$ and the limit 
\be\la{2.45}
R_- := \slim_{t\to-\infty}U(t)^*(H_D + \gt)^{-1}U(t)
\ee
exist.  Moreover, 
\be\label{ranperp}
{\rm Ran }^\perp(\gO_-)={\rm Ker}(R_-).
\ee
\ep
\begin{proof} We start with \eqref{2.45}. 
Let us introduce the time-dependent identification operator
\bed
J_D(t) := (H_\ga(t) + \gt)^{-1}(H_D + \gt)^{-1}, \quad t \in \bR.
\eed
We have
\bea\la{2.47}
\lefteqn{
\frac{d}{dt}U(t)^*J_D(t)e^{-itH_D}f =}\\
& &
iU(t)^*\left((H_D + \gt)^{-1} - (H_\ga(t) + \gt)^{-1}\right)e^{-itH_D}f 
+ U(t)^*\dot{J}_D(t)e^{-itH}f
\nonumber
\eea
for $f \in \gotH$ where $\dot{J}_D := \frac{d}{dt}J_D(t)$. 
Hence we get
\bea\la{3.18}
\lefteqn{
U(t)J_D(t)e^{-itH_D}f - U(s)^*J_D(s)e^{-isH_D}f =}\\
& &
i\int^t_s U(s)^*\left((H_D + \gt)^{-1} - (H_\ga(r) +
  \gt)^{-1}\right)e^{-irH_D}f dr +
\nonumber \\
& &
\int^t_0 U(r)^*\dot{J}_D(r)e^{-irH_D}f ds.
\nonumber
\eea
Using \eqref{2.35} we find
\bea\label{una1}
\lefteqn{
(H_\ga(t) + \gt)^{-1} = (H + \gt)^{-1/2}Q_B(H + \gt)^{-1/2} +}\\
& &
e^{\ga t}(H + \gt)^{-1/2}Q^\perp_B(e^{\ga t} + B)^{-1}Q^\perp_B(H + \gt)^{-1/2}
\nonumber
\eea
where $Q_B$ is the orthogonal projection onto the subspace ${\rm
  Ker}(B)$. Note that $Q_B^\perp$ has rank 2. 
Taking into account \eqref{1.31} we get the representation
\be\la{2.49}
(H_D + \gt)^{-1} = (H + \gt)^{-1/2}Q_B(H + \gt)^{-1/2}.
\ee
By \eqref{una1}  and \eqref{2.49} we obtain
\bea\la{3.21}
\lefteqn{
(H_D + \gt)^{-1} - (H_\ga(t) + \gt)^{-1} =}\\
& &
-e^{\ga t}(H + \gt)^{-1/2}Q^\perp_B(e^{\ga t} + B)^{-1}Q^\perp_B(H + \gt)^{-1/2}.
\nonumber
\eea
Since $B$ is positive and invertible on ${\rm Ker}^\perp(B)$ we get the estimate
\be\la{2.51}
\|(H_D + \gt)^{-1} - (H_\ga(t) + \gt)^{-1}\| \le
e^{\ga t}\|Q^\perp_BB^{-1}Q^\perp_B\|,
\quad t \in \dR, \quad \ga > 0.
\ee
Using again \eqref{una1} we have 

\be\la{4.5}
\dot{J}_D(t) = \ga e^{\ga t}(H + \gt)^{-1/2}Q_B^\perp(e^{\ga t} + B)^{-2}BQ_B^\perp(H + \gt)^{-1/2}(H_D + \gt)^{-1}.
\ee
This gives the estimate
\be\la{2.54}
\|\dot{J}_D(t)\| \le \ga e^{\ga t}\|Q^\perp_B B^{-1}Q^\perp_B\|.
\ee
Using \eqref{3.18}, \eqref{2.51} and \eqref{2.54} we prove the
existence of the limit
\bed
\widehat{\gO}_- := \slim_{t\to-\infty}U(t)^*J_D(t)e^{-itH_D}.
\eed
In fact, the convergence is in operator norm:
\be\la{2.56}
\lim_{t\to-\infty}\|\widehat{\gO}_- - U(t)^*J_D(t)e^{-itH_D}\| = 0.
\ee
Using the identity
\bead
\lefteqn{
U(t)^*J_D(t)e^{-itH_D} - U(t)^*e^{-itH_D}(H_D + \gt)^{-2} }\\
& &
=U(t)^*((H_\ga(t) + \gt)^{-1} - (H_D +
\gt)^{-1})e^{-itH_D}(H_D + \gt)^{-1} 
\nonumber
\eead
and \eqref{2.51} we get the estimate
\be\la{2.57} 
\|U(t)^*J_D(t)e^{-itH_D} - U(t)^*e^{-itH_D}(H_D + \gt)^{-2}\| 
\le e^{\ga t}\|Q^\perp_BB^{-1}Q^\perp_B\|,
\ee
which yields
\be\la{2.58}
\lim_{t\to-\infty}\|\widehat{\gO}_- - U(t)^*e^{-itH_D}(H_D + \gt)^{-2}\| = 0,
\ee
for all $t \in \dR$, $\ga > 0$. Since the wave operator $\widehat{\gO}_-$ exists we
get the existence of 
$$\slim_{t\to-\infty}U(t)^*e^{-itH_D}(H_D + \gt)^{-2}.$$
Using that $\ran((H_D + \gt)^{-2})$ is dense in $\gotH$, we prove the
existence of $\gO_-$. In particular, this proves that $\gO_-$ is
isometric, i.e $\gO^*_-\gO_- = I$.

Now let us prove that the operator in \eqref{2.45} exists. Note that
the norm convergence in \eqref{2.56} yields the same property for adjoints:
\bed
\lim_{t\to-\infty}\|\widehat{\gO}^*_- - e^{itH_D}J_D(t)^*U(t)\| = 0.
\eed
In particular
\bed
\widehat{\gO}^*_- = \slim_{t\to-\infty}e^{itH_D}J_D(t)^*U(t).
\eed
In the quadratic form sense we get
\bed
\frac{d}{dt}U(t)^*(H_\ga(t) + \gt)^{-1}U(t)f =
U(t)^*\left\{\frac{d}{dt}(H_\ga(t) + \gt)^{-1}\right\}U(t)f
\eed
$f \in \gotH$, $t \in \dR$, $\ga > 0$. Hence
\begin{align}
U(t)^*(H_\ga(t) + \gt)^{-1}&U(t)f -  U(s)^*(H_\ga(t) + \gt)^{-1}U(s)f 
\nonumber \\
& =\int^t_s dr\; U(r)^*\left\{\frac{d}{dr}(H_\ga(r) +\gt)^{-1}\right\}U(r)f
\nonumber
\end{align}
$f \in \gotH$, $t,s \in \dR$, $\ga > 0$.
By \eqref{2.35} we get
\bed
\frac{d}{dt}(H_\ga(t) + \gt)^{-1} = 
\ga e^{-\ga t} (H + \gt)^{-1/2}(I + e^{-\ga t}B)^{-2}B(H + \gt)^{-1/2}
\eed
which gives the estimate
\be\label{derivrezo}
\left\|\frac{d}{dt}(H_\ga(t) + \gt)^{-1}\right\|
\le \ga e^{\ga t}\|Q^\perp_B B^{-1} Q^\perp_B\|.
\ee
Hence $R_-$ exists, and we even have convergence in operator norm:
\bed
\lim_{t\to-\infty}\|R_- - U(t)^*(H_\ga(t) + \gt)^{-1}U(t)^*\| = 0.
\eed
Taking into account the estimate \eqref{2.51} we find
\be\label{amnevoie}
\lim_{t\to-\infty}\|R_- - U(t)^*(H_D + \gt)^{-1}U(t)^*\| = 0.
\ee
In particular we have
\bead
\lefteqn{
\lim_{t\to-\infty}\|R^2_- - U(t)^*(H_\ga(t) + \gt)^{-2}U(t)^*\| }\\
& &
=\lim_{t\to-\infty}\|R^2_- - U(t)^*(H_D + \gt)^{-2}U(t)^*\| = 0.
\nonumber
\eead
Using the identity
\bed
J_D(t)^* = \left((H_D + \gt)^{-1} - (H_\ga(t) +
  \gt)^{-1}\right)(H_\ga(t) + \gt)^{-1}
+ (H_\ga(t) + \gt)^{-2}
\eed
and taking into account the estimate \eqref{2.51} we obtain
\bed
\lim_{t\to-\infty}\|R^2_- -  U(t)^*J_D(t)^*U(t)\| = 0.
\eed
Hence we find
\bea\la{3.43}
\lefteqn{
\widehat{\gO}^*_- = 
\slim_{t\to-\infty}e^{itH_D}J_D(t)^*U(t)}\\
& &
=\slim_{t\to-\infty}e^{itH_D}U(t)U(t)^*J_D(t)^*U(t) =
\slim_{t\to-\infty}e^{itH_D}U(t)R^2_-
\nonumber
\eea
which shows in particular that the limit $\lim_{t\to-\infty}e^{itH_D}U(t)f$ exist
for elements $f \in \overline{\ran(R_-)}$. More precisely:
\be\label{convperan}
\lim_{t\to-\infty}e^{itH_D}U(t)R_-f=\gO^*_-R_-f
\ee
for all $f$.

\vspace{0.5cm} We are now ready to prove \eqref{ranperp}. 
Assume that $f \perp \ran(\gO_-)$. Then using the definitions, the
unitarity of $U(t)^*$, and \eqref{amnevoie} we obtain:
\bead
\lefteqn{
0 = (f,\gO_-(H_D + \gt)^{-1}g) }\\
& &
=\lim_{t\to-\infty}(U(t)^*(H_D + \gt)^{-1}U(t)f,U(t)^*e^{-itH_D}g) =
(R_-f,\gO_-g) 
\nonumber
\eead
for $g \in \gotH$.  Hence $f \perp \ran(\gO_-)$ implies $R_-f \perp
\ran(\gO_-)={\rm Ker}(\gO_-^*)$.  Thus  
$\gO^*_-R_-f = 0$. Using \eqref{convperan}:
\bed
0 = \lim_{t\to-\infty}\|e^{itH_D}U(t)R_-f\| = \|R_-f\|, 
\eed
thus $f \in {\rm Ker}(R_-)$. We have thus shown that
$\ran^\perp(\gO_-)\subset{\rm Ker}(R_-)$. Conversely, choose $f \in
{\rm Ker}(R_-)$. We have (use \eqref{amnevoie}):
\bead
\lefteqn{
(f,\gO_-(H_D + \gt)^{-1}g) = 
\lim_{t\to-\infty}(f,U(t)^*e^{-itH_D}(H_D + \gt)^{-1}g)
}\\
& &
=\lim_{t\to-\infty}(U(t)^*(H_D + \gt)^{-1}U(t)f,U(t)^*e^{-itH_D}g) =
(R_-f,\gO_-g)=0,
\nonumber
\eead
for all $g$. Thus $\gO_-^*f$ is orthogonal on a dense set (domain of
$H_D$), thus equals zero. Therefore ${\rm Ker}(R_-)\subset
\ran^\perp(\gO_-)$ and \eqref{ranperp} is proved.

\noindent {\bf Remark}. Note that $\gO_-$ would be unitary if one
could prove that ${\rm Ker}(R_-)=\emptyset$. 

\end{proof}
\bp\la{II.2}
Let $H$ and $H_\ga(t)$, $t \in \dR$, $\ga > 0$, be given by 
\eqref{1.1}-\eqref{1.4} and \eqref{1.29}-\eqref{1.30}, respectively.
Then the wave operator $\gO_+$ exists and is unitary.
\ep
\begin{proof}
We introduce the identification operator
\bed
J(t) := (H_\ga(t) + \gt)^{-1}(H + \gt)^{-1}.
\quad t \in \dR.
\eed
In the quadratic form sense we get that
\bea\la{3.49}
\lefteqn{
\frac{d}{dt}U(t)^*J(t)e^{-itH}f =}\\
& &
iU(t)^*((H + \gt)^{-1} - (H_\ga(t) + \gt)^{-1})e^{-itH}f +
U(t)^*\dot{J}(t)e^{-itH}f,
\nonumber
\eea
$t \in \bR$, where
\bed
\dot{J}(t) := \frac{d}{dt}J(t).
\eed
Taking into account \eqref{2.35} we find
\be\label{diferenres}
(H + \gt)^{-1} - (H_\ga(t) + \gt)^{-1} = e^{-\ga t}(H + \gt)^{-1/2}(I +
e^{-\ga t}B)^{-1}B(H + \gt)^{-1/2}, 
\ee
$t \in \dR$. Hence we have the estimate
\be\la{3.52}
\|(H + \gt)^{-1} - (H_\ga(t) + \gt)^{-1}\| \le e^{-\ga t}\|B\|, \quad
t \in \dR.
\ee
Moreover, we get
\bea\la{3.53}
\lefteqn{
\dot{J}(t) = \frac{d}{dr}(H_\ga(t) + \gt)^{-1}(H + \gt)^{-1}}\\
& & 
=\ga e^{-\ga t}(H + \gt)^{-1/2}(I + e^{-\ga t}B)^{-2}B(H + \gt)^{-3/2}
\nonumber
\eea
which yields the estimate 
\bed
\left\|\dot{J}(t)\right\| \le \ga e^{-\ga t}\|B\|, 
\quad t \in \dR.
\eed
Hence the strong limit
\bed
\widehat{\gO}_+ := \slim_{t\to+\infty}U(t)^*J(t)e^{-itH} 
\eed
exists. Moreover, the convergence is also true in operator norm:
\be\la{3.56}
\lim_{t\to+\infty}\|\widehat{\gO}_+ - U(t)^*J(t)e^{-itH}\| = 0.
\ee
Using the identity
\bea\la{3.57}
\lefteqn{
U(t)^*J(t)e^{-itH} - U(t)^*e^{-itH}(H + \gt)^{-2} 
= }\\
& &
U(t)^*\left((H_\ga(t) + \gt)^{-1} - (H + \gt)^{-1}\right)e^{-itH}(H + \gt)^{-1}
\nonumber
\eea
and taking into account the estimate \eqref{3.52} we obtain
\be\la{3.58}
\lim_{t\to+\infty}\|\widehat{\gO}_+ - U(t)^*e^{-itH}(H + \gt)^{-2}\| =
0.
\ee
Hence $\gO_+$ exists on a dense domain and is isometric, i.e. $\gO^*_+\gO_+ =
I$. 

Let us now prove that $\gO_+$ is unitary. Since \eqref{3.56} holds
also true for adjoints we
find
\be\la{3.59}
\lim_{t\to+\infty}\|\widehat{\gO}^*_+ - e^{itH}J(t)U(t)\| = 0.
\ee
Hence, we have the representation
\be\label{omegahatplus}
\widehat{\gO}^*_+ = \slim_{t\to+\infty}e^{itH}J(t)U(t).
\ee
Furthermore, in the quadratic form sense we have
\bead
\lefteqn{
\frac{d}{dt} U(t)^*(H_\ga(t) + \gt)^{-1}U(t)
=U(t)^*\left\{\frac{d}{dt}(H_\ga(t) + \gt)^{-1}\right\}U(t) =}\\
& &
\ga e^{-\ga t}U(t)^*(H + \gt)^{-1/2}(I + e^{-\ga t}B)^{-2}B(H +
\gt)^{-1/2}U(t).
\nonumber 
\eead
Hence we get
\bea\la{3.62}
\lefteqn{
U(t)^*(H_\ga(t) + \gt)^{-1}U(t)= (H_\ga(0) + \gt)^{-1}}\\
& &
+\ga\int^t_0 \;e^{-\ga t}U(t)^*(H + \gt)^{-1/2}(I + e^{-\ga t}B)^{-2}B(H +
\gt)^{-1/2}U(t)dt.
\nonumber
\eea
Using the estimate
\bed
\left\|U(t)^*(H + \gt)^{-1/2}(I + e^{-\ga t}B)^{-2}B(H +
\gt)^{-1/2}U(t)\right\| \le
\|B\|, \quad t \in \dR,
\eed
we find that the following weak integral exists and defines a bounded operator:
\be\label{operadoi}
\ga\int^\infty_0 dt\;e^{-\ga t}U(t)^*(H + \gt)^{-1/2}(I + e^{-\ga t}B)^{-2}B(H +
\gt)^{-1/2}U(t).
\ee
Moreover, by the Cook argument it also implies the existence of the limit
\bed
R_+ := \slim_{t\to+\infty}U(t)^*(H_\ga(t) + \gt)^{-1}U(t).
\eed
In fact, the convergence takes place in operator norm:
\bed
\lim_{t\to+\infty}\|R_+ - U(t)^*(H_\ga(t) + \gt)^{-1}U(t)\| = 0.
\eed
Taking into account the estimate \eqref{3.52} we obtain
\bed
\lim_{t\to+\infty}\|R_+ - U(t)^*(H + \gt)^{-1}U(t)\| = 0.
\eed
which yields 
\bed
\lim_{t\to+\infty}\|R^2_+ - U(t)^*J(t)U(t)\| = 0.
\eed
By \eqref{3.59} we get
\bed
\lim_{t\to+\infty}\|\widehat{\gO}^*_+ - e^{itH}U(t)R^2_+\| = 0
\eed
which shows the existence of $\gO^*_+f =
\slim_{t\to+\infty}e^{itH}U(t)f$ for $f \in \overline{\ran(R_+)}$. Now
in order to prove the unitarity of $\gO_+$ it is enough to show that
$\ran(R_+)$ is dense in $\gotH$. Let us do that. 

From \eqref{3.62} we obtain
\bead
\lefteqn{
R_+ = (H_\ga(0) + \gt)^{-1} + }\\
& &
\ga\int^\infty_0 dt\;e^{-\ga t}U(t)^*(H + \gt)^{-1/2} (I + e^{-\ga t}B)^{-2}B(H + \gt)^{-1/2}U(t),
\nonumber
\eead
which by the positivity of the integral it gives $0 \le (H_\ga(0) +
\gt)^{-1} \le R_+$. Hence 
\bed
{\rm Ker}(R^{1/2}_+)\subseteq {\rm Ker}((H_\ga(0) + \gt)^{-1/2}),
\eed
thus
\bed
\ran^\perp(R^{1/2}_+) \subseteq\ran^\perp ((H_\ga(0) + \gt)^{-1/2})=\emptyset.
\eed
Thus we get that
$\ran(R^{1/2}_+)$ is dense in $\gotH$ which yields that $\ran(R^2_+)$
is dense in $\gotH$. Therefore we have the representation $\gO^*_+ =
\slim_{t\to+\infty}e^{itH}U(t)$ which proves that $\gO_+$ is unitary.
\end{proof}

\subsection{Time-dependent density operator}

Now we are ready to write down a solution to our Liouville equation
\eqref{1.33} which also obeys the initial condition at
$t=-\infty$. Let us introduce the notation:
\bed
\gr_\ga(0) := \gO_-\gr_D\gO^*_-
\eed
which defines a non-negative self-adjoint operator. Here $\gr_D$
is given by \eqref{1.25}-\eqref{1.27}. In accordance with \eqref{3.9},
the time evolution of $\gr_\ga(0)$ is given by
\be\label{soluliu}
\gr_\ga(t) = U(t)\gr_\ga(0)U(t)^* = U(t)\gO_-\gr_D\gO^*_-U(t)^*,
\quad t \in \dR,
\ee
where we have used the notation $U(t) := U(t,0)$ and the relation
$U(0,t) = U(t)^*$, $t \in \dR$. We now show that the initial condition
is fulfilled.
\bp\label{propo3.3}
Let $H_D$ and $H_\ga(t)$, $t \in \dR$, $\ga > 0$, be given by
\eqref{1.16}-\eqref{1.24} and \eqref{1.29}-\eqref{1.30}, respectively.
If $\gr_D$ is a steady state for the system $\{\gotH,H_D\}$ such that 
the operator $\widehat{\gr}_D := (H_D + \gt)^4\gr_D$ is bounded, then
\be\la{4.3}
\lim_{t\to-\infty}\|\gr_D - \gr_\ga(t)\| = 0.
\ee
\ep
\begin{proof}
We write the identity:
\be\la{4.4}
 U(t)\gO_-\gr_D\gO^*_-U(t)^* = 
U(t)\gO_-(H_D + \gt)^{-2}\widehat{\gr}_D(H_D + \gt)^{-2}\gO^*_-U(t)^*,
\ee
$t \in \dR$. Taking into account \eqref{2.58} we find
\be\la{4.5bis}
 U(t)\gO_-\gr_D\gO^*_-U(t)^* = 
U(t)\widehat{\gO}_-\widehat{\gr}_D\widehat{\gO}^*_-U(t)^*.
\ee
From \eqref{2.56} we get
\begin{align}\la{4.6}
\lim_{t\to-\infty}\|U(t)\widehat{\gO}_-\widehat{\gr}_D\widehat{\gO}^*_-U(t)^*
- J_D(t)e^{-itH_D}&\widehat{\gr}_De^{itH_D}J_D(t)^*\| \\
=\lim_{t\to-\infty}\|U(t)\widehat{\gO}_-\widehat{\gr}_D\widehat{\gO}^*_-U(t)^*&
- J_D(t)\widehat{\gr}_DJ_D(t)^*\| = 
0.
\nonumber
\end{align}
Using \eqref{2.51} we get
\begin{align}\la{4.7}
\lim_{t\to-\infty}\|J_D(t)\widehat{\gr}_DJ_D(t)^* - 
(H_D + &\gt)^{-2}\widehat{\gr}_D(H_D + \gt)^{-2}\|  \\
&=\lim_{t\to-\infty}\|J_D(t)\widehat{\gr}_DJ_D(t)^* - \gr_D\| =
0
\nonumber
\end{align}
Taking into account \eqref{4.4}-\eqref{4.7} we prove \eqref{4.3}.
\end{proof}

\subsection{Large time behavior on the space of absolute continuity}
We now are ready to prove the result announced in \eqref{surprise}.
\bp\la{III.4}
Let $H$ and $H_\ga(t)$, $t \in \dR$, $\ga > 0$, be given by 
\eqref{1.1}-\eqref{1.4} and \eqref{1.29}-\eqref{1.30},
respectively. Let $W_-$ be the incoming wave operator as defined in
\eqref{incwaveop}. 
If $\gr_D$ is a steady state for the system $\{\gotH,H_D\}$ such that 
the operator $\widehat{\gr}_D := (H_D + \gt)^4\gr_D$ is bounded, then
\be\la{4.8}
\slim_{t\to+\infty}\gr_\ga(t)P^{ac}(H)=W_- \gr_D W^*_- .
\ee
\ep
\begin{proof}

Let us assume that the following three technical results hold true:
\be\la{4.20}
\slim_{t\to+\infty}(U(t)^* - e^{itH})P^{ac}(H) = 0,
\ee
\be\label{4.20bis}
(H_D + \gt)^{-2}(\gO^*_- -  I)\quad {\rm is}\quad {\rm compact},
\ee
and 
\be\la{4.24}
\slim_{t\to+\infty}(H_D + \gt)^{-2}(\gO^*_- - I)e^{itH}P^{ac}(H) = 0.
\ee

We will first use these estimates in order to prove the proposition,
and then we will give their own proof.

\vspace{0.5cm}

We write the identity:
\bead
\lefteqn{
U(t)\gr_\ga(t)U(t)^*P^{ac}(H)}\\
& & 
=U(t)\gO_-(H_D + \gt)^{-2}\widehat{\gr}_D(H_D + \gt)^{-2}\gO^*_-U(t)^*P^{ac}(H) \nonumber\\
& &
=U(t)\gO_-(H_D + \gt)^{-2}\widehat{\gr}_D(H_D + \gt)^{-2}\gO^*_-(U(t)^*
- e^{itH})P^{ac}(H) \nonumber\\
& &
+U(t)\gO_-(H_D + \gt)^{-2}\widehat{\gr}_D(H_D + \gt)^{-2}(\gO^*_- - I)
e^{itH}P^{ac}(H)\nonumber\\
& &
+U(t)\gO_-(H_D + \gt)^{-2}e^{itH_D}\widehat{\gr}_D(H_D + \gt)^{-2}e^{-itH_D}e^{itH}P^{ac}(H).
\nonumber
\eead
Taking into account \eqref{4.20}-\eqref{4.24}, and using the completeness of $W_-$ which yields $W^*_- =
\slim_{t\to-\infty}e^{itH_D}e^{-itH}P^{ac}(H)$, we get:
\bead
\lefteqn{
\slim_{t\to+\infty}U(t)\gr_\ga(t)U(t)^*P^{ac}(H) }\\
& &
=\slim_{t\to+\infty}U(t)\gO_-(H_D + \gt)^{-2}e^{itH_D}\widehat{\gr}_D(H_D + \gt)^{-2}W^*_-.
\nonumber
\eead
Since $(\gO_- - I)(H_D + \gt)^{-2}$ is also compact (its adjoint is
compact, see \eqref{4.20bis}), we have:
\bed
\slim_{t\to+\infty}(\gO_- - I)(H_D + \gt)^{-2}e^{itH_D}P^{ac}(H_D) =
0.
\eed
Thus:
\bead
\lefteqn{
\slim_{t\to+\infty}U(t)\gr_\ga(t)U(t)^*P^{ac}(H) =}\\
& &
\slim_{t\to+\infty}U(t)e^{itH_D}(H_D + \gt)^{-2}\widehat{\gr}_D(H_D + \gt)^{-2}W^*_- 
\nonumber\\
& &
= \slim_{t\to+\infty}U(t)e^{itH_D}\gr_D W^*_- =
\slim_{t\to+\infty}U(t)e^{itH}P^{ac}(H)W_-\gr_D W^*_-.
\nonumber
\eead
Finally, we apply \eqref{4.20} once again, and \eqref{4.8} is proved.

\vspace{0.5cm}

Now let us prove the three technical results announced in
\eqref{4.20}-\eqref{4.24}. We start with \eqref{4.20}.

We have the identity:
\be\label{adoua1}
(U(t)^* - e^{itH})(H + \gt)^{-2} = 
(U(t)^*(H + \gt)^{-2}e^{-itH} - (H+\gt)^{-2})e^{itH}.
\ee
Then by adding and subtracting several terms we can write another identity:
\bea
\lefteqn{
(U(t)^* - e^{itH})(H + \gt)^{-2}\nonumber}\\
& &
=\left\{U(t)^*(H + \gt)^{-2}e^{-itH} - \widehat{\gO}_+\right\}e^{itH} \la{4.12}\\
& & 
+\left\{\widehat{\gO}_+ - J(0)\right\}e^{itH}g + \left\{J(0) - (H+\gt)^{-2}\right\}e^{itH}.
\la{4.13}
\eea
By \eqref{3.58} we get
\begin{align}\label{4.13a}
\lim_{t\to+\infty}\|U(t)^*(H + \gt)^{-2}e^{-itH} -
\widehat{\gO}_+\| = 0.
\end{align}
which shows that \eqref{4.12} tends to zero as $t \to + \infty$.
Next, from \eqref{2.35}, \eqref{3.49} and \eqref{3.53} we get
\bead
\lefteqn{
U(t)^*J(t)e^{-itH} - J(0) }\\
& &
=i\int^t_0 ds \; e^{-\ga s}U(s)^*(H + \gt)^{-1/2}(I + e^{-\ga t}B)^{-1}B(H + \gt)^{-1/2}e^{-isH} 
\nonumber\\
& &
+\ga\int^t_0 ds\;e^{-\ga s} U(s)^*(H + \gt)^{-1/2}(I + e^{-\ga s}B)^{-2}B(H + \gt)^{-3/2}e^{-isH},
\nonumber
\eead
which yields
\bead
\lefteqn{
\widehat{\gO}_+ - J(0)}\\
& &
=i\int^\infty_0 ds \; e^{-\ga s}U(s)^*(H + \gt)^{-1/2}
(I + e^{-\ga s}B)^{-1}B(H + \gt)^{-1/2}e^{-isH} 
 \nonumber\\
& &
+\ga\int^\infty_0 ds\;e^{-\ga s} U(s)^*(H + \gt)^{-1/2}
(I + e^{-\ga s}B)^{-2}B(H + \gt)^{-3/2}e^{-isH}.
\nonumber
\eead
Since $B$ is a compact (rank 2) operator, 
we get that $\widehat{\gO}_+ - J(0)$ is a compact operator. This fact
immediately implies (via the RAGE theorem):
\be\la{4.16}
\slim_{t\to+\infty}(\widehat{\gO}_+ - J(0))e^{itH}P^{ac}(H) = 0.
\ee
Furthermore, we have the identity:
\bead
\lefteqn{
J(0) - (H + \gt)^{-2} }\\
& &
=\left((H_\ga(0) + \gt)^{-1} - (H + \gt)^{-1}\right)(H + \gt)^{-1} 
\nonumber\\
& &
=-(H + \gt)^{-1/2}(I + B)^{-1}B(H + \gt)^{-3/2},
\nonumber
\eead
which gives that $J(0) - (H + \gt)^{-2}$ is compact. Thus:
\be\la{4.18}
\slim_{t\to+\infty}(J(0) - (H + \gt)^{-2})e^{itH}P^{ac}(H) = 0.
\ee
Taking into account \eqref{4.13a}, \eqref{4.16} and \eqref{4.18}
we find
\bed
\slim_{t\to+\infty}(U(t)^* - e^{itH})(H + \gt)^{-2}P^{ac}(H) = 0
\eed
which proves \eqref{4.20}. 

Next we prove \eqref{4.20bis} (the estimate \eqref{4.24} is just an easy
consequence of \eqref{4.20bis} via the RAGE theorem). Using \eqref{2.58} we have
$\gO_-(H_D + I)^{-2} = \widehat{\gO}_-$. From \eqref{3.18},
\eqref{3.21} and \eqref{4.5} we obtain:
\bead
\lefteqn{
J_D(0) - U(t)^*J_D(t)e^{-itH_D} }\\
& &
=-i\int^0_t ds\;e^{\ga s} U(s)^*
(H + \gt)^{-1/2}Q^\perp_B(e^{\ga s} + B)^{-1}Q^\perp_B(H + \gt)^{-1/2}e^{-isH_D} 
\nonumber\\
& &
+\ga \int^0_t ds\; e^{\ga s} U(s)^*
(H + \gt)^{-1/2}(e^{\ga s} + B)^{-2}B(H + \gt)^{-1/2}(H_D + \gt)^{-1}
e^{-isH_D}
\nonumber
\eead
and together with the fact that $Q^\perp_B$ is a rank 2 operator 
we find that $\widehat{\gO}_- - J_D(0)$ is compact.
Hence $\widehat{\gO}^*_- - J_D(0)^*$ is compact, too.

Moreover, using \eqref{3.21} we get
\bead
\lefteqn{
J_D(0) - (H_D + \gt)^{-2} = }\\
& & 
((H_\ga(0) + \gt)^{-1} - (H_D + \gt)^{-1})(H_D + \gt)^{-1}
= \nonumber\\
& &
-e^{\ga t}(H + \gt)^{-1/2}Q^\perp_B(e^{\ga t} + B)^{-1}Q^\perp_B(H + \gt)^{-1/2}(H_D + \gt)^{-1}
\nonumber
\eead
which shows that $J_D(0) - (H_D + \gt)^{-2}$ is compact. Hence
$J_D(0)^* - (H_D + \gt)^{-2}$ is compact.

Now use the identity:
\bed
(H_D + \gt)^{-2}(\gO^*_- -  I) = (\widehat{\gO}^*_- - J_D(0)) + (J_D(0) -
(H_D + \gt)^{-2}), 
\eed
which proves \eqref{4.20bis}. Finally, \eqref{4.24} follows from
\eqref{4.20bis} and the RAGE theorem. 
\end{proof}

\subsection{The main result}

We are now ready to rigorously formulate and prove our main result,
announced in the introduction:

\bt \label{maintheorem}
Let $H$ and $H_\ga(t)$, $t \in \dR$, $\ga > 0$, be given by 
\eqref{1.1}-\eqref{1.4} and \eqref{1.29}-\eqref{1.30},
respectively. Let $W_-$ be the incoming wave operator from
\eqref{incwaveop}. 
Further, let $E_H(\cdot)$ and $\{\gl_j\}^N_{j=1}$ be the spectral
measure and the eigenvalues of $H$. 
If $\gr_D$ is a steady state for the system $\{\gotH,H_D\}$ such that 
the operator $\widehat{\gr}_D := (H_D + \gt)^4\gr_D$ is bounded, then
the limit 
\bea\la{3.100}
\lefteqn{
\gr_\ga := \slim_{T\to+\infty}\frac{1}{T}\int^T_0 dt \gr_\ga(t) =}\\
& &
\sum^N_{j=1} E_H(\{\gl_j\})S_\ga \gr_D S^*_\ga E_H(\{\gl_j\}) +
W_-\gr_D W^*_-
\nonumber
\eea
exists and defines a steady state for the system $\{\gotH,H\}$ where
$S_\ga := \gO^*_+\gO_-$.
\et
\br
{\em
We stress once again that only the part corresponding to
the pure point spectrum $\gr^p_\ga := \sum^N_{j=1}E_H(\{\gL_j\})S_\ga \gr_D
S^*_\ga E_H(\{\gl_j\})$ of the steady state $\gr_\ga$
depends on the parameter $\ga > 0$
while the absolutely continuous part $\gr^{ac}_\ga :=
W_-\gr_DW^*_-$ does not. Note that with respect to the
decomposition $\gotH = \gotH^p(H) \oplus \gotH^{ac}(H)$, one has $\gr_\ga = \gr^p_\ga
\oplus \gr^{ac}_\ga$.
}
\er
\begin{proof}
By Proposition \ref{III.4} we have
\be\la{3.101}
\slim_{T\to+\infty}\frac{1}{T}\int^T_0 dt \; \gr_\ga(t)P^{ac}(H) = W_-\gr_DW^*_-.
\ee
In particular, this yields:
\bed
\slim_{T\to+\infty}\frac{1}{T}\int^T_0 dt \; P^s(H)\gr_\ga(t)P^{ac}(H) = 0,
\eed
where $P^s(H)$ is the projection onto the singular subspace of $H$. 
Now we are going to prove 
\be\la{3.102}
\slim_{T\to+\infty}
\frac{1}{T}\int^T_0 P^{ac}(H)\gr_\ga(t)P^{s}(H) dt = 0.
\ee
By \eqref{3.58} we find
\bed
\lim_{t\to\infty}
\|U(t)\gO_-\gr_D\gO^*_-U(t)^*(I + H)^{-2}e^{-itH} - 
U(t)\gO_-\gr_D\gO_-^*\widehat{\gO}_+\| = 0,
\eed
which yields
\bed
\lim_{t\to\infty}
\|U(t)\gO_-\gr_D\gO^*_-U(t)^*(I + H)^{-2} -
U(t)\gO_-\gr_D\gO_-^*\widehat{\gO}_+e^{itH}\| = 0.
\eed
Let $\gl_j$ be an eigenvalue of $H$ with corresponding to an eigenfunction $\phi_j$. Then
\bed
\lim_{t\to\infty}
\|U(t)\gO_-\gr_D\gO^*_-U(t)^*(I + H)^{-2}\phi_j -
e^{it\gl_j}U(t)\gO_-\gr_D\gO_-^*\widehat{\gO}_+\phi_j\| = 0.
\eed
This and the unitarity of $\gO^*_+$ give:
\bed
\lim_{t\to\infty}
\|\gr_\ga(t)(I + H)^{-2}\phi_j -
e^{it\gl_j}e^{-itH}\gO^*_+\gO_-\gr_D\gO_-^*\gO_+(H + \gt)^{-2}\phi_j\| = 0
\eed
or
\be\la{3.108}
\lim_{t\to\infty}
\|\gr_\ga(t)\phi_j -
e^{it\gl_j}e^{-itH}\gO^*_+\gO_-\gr_D\gO_-^*\gO_+\phi_j\| = 0.
\ee
Hence we have
\bed
\frac{1}{T} \int^T_0 dt\; P^{ac}(H)\gr_\ga(t)\phi_j =
\frac{1}{T} \int^T_0 dt\;e^{it\gl_j}e^{-itH}P^{ac}(H)\gO^*_+\gO_-\gr_D\gO_-^*\gO_+\phi_j.
\eed
We use the decomposition
\bead
\lefteqn{
\frac{1}{T}\int^T_0 P^{ac}(H)\gr_\ga(t)\phi_j }\\
& &
=\frac{1}{T}\int^T_0 dt\; e^{it\gl_j}e^{-itH}E_H(|\gl - \gl_j| < \epsilon)P^{ac}(H)
\gO^*_+\gO_-\gr_D\gO_-^*\gO_+\phi_j
\nonumber\\
& &
+\frac{1}{T}\int^T_0 dt \; e^{it\gl_j}e^{-itH}E_H(|\gl - \gl_j| \ge \epsilon)P^{ac}(H)
\gO^*_+\gO_-\gr_D\gO_-^*\gO_+\phi_j.
\nonumber
\eead
If $\epsilon$ is small enough, then $E_H(|\gl - \gl_j| < \epsilon)P^{ac}(H)=0$. 
This yields the estimate:
\bead
\lefteqn{
\left\|\frac{1}{T}\int^T_0 P^{ac}(H)\gr_\ga(t)\phi_j\right\|}\\
& &
\leq \frac{2}{T}\left\|
(H - \gl_j)^{-1}E_H(|\gl - \gl_j| \ge \epsilon)P^{ac}(H)
\gO^*_+\gO_-\gr_D\gO_-^*\gO_+\phi_j
\right\|
\nonumber
\eead
which immediately shows that
\be\la{3.112}
\slim_{T\to\infty}\frac{1}{T}\int^T_0 dt \; P^{ac}(H)\gr_\ga(t)\phi_j
= 0,
\ee
and \eqref{3.102} is proved. Next, from \eqref{3.108} we easily obtain:
\be\la{3.113}
\slim_{T\to\infty}\frac{1}{T}\int^T_0 dt\;P^{s}(H)\gr_\ga(t)\phi_j =
 E_H(\{\gl_j\})\,S_\ga\,\gr_D\,S^*_\ga E_H(\{\gl_j\}).
\ee
Now put together \eqref{3.101}, \eqref{3.102}, \eqref{3.112} and
\eqref{3.113}, and the proof of \eqref{3.100} is over. Now the operator $\gr_\ga$ is
non-negative, bounded, and commutes with $H$. Hence $\gr_\ga$ is a
steady state for $\{\gotH,H\}$. 
\end{proof}
\bc
Let $H$ and $H_\ga(t)$, $t \in \dR$, $\ga > 0$, be given by 
\eqref{1.1}-\eqref{1.4} and \eqref{1.29}-\eqref{1.30}, respectively.
Then with respect to the spectral representation
$\{L^2(\dR,\goth(\gl),\nu),M\}$ of $H$ the distribution function
$\{\tilde{\rho}_\ga(\gl)\}_{\gl \in \dR}$ of the steady state $\gr_\ga$ is given by
\be\la{3.114}
\tilde{\rho}_\ga(\gl) := 
\begin{cases}
0, & \gl \in \dR \setminus \gs(H)\\
\rho_{\ga,j}, & \gl = \gl_j, \quad j = 1,\ldots,N\\
\gotf_b(\gl-\mu_b), & \gl \in [v_b,v_a)\\
\begin{pmatrix}
\gotf_b(\gl-\mu_b) & 0\\
0 & \gotf_a(\gl -\mu_a)
\end{pmatrix}, & \gl \in [v_a,\infty)
\end{cases}
\ee
where $\rho_{\ga,j} := (S_\ga\phi_j,\phi_j)$, $j =
1,2,\ldots,N$.
\ec
\begin{proof}
Using the generalized Fourier transform \eqref{2.23}-\eqref{2.25} one
has to consider the operator $\Phi^{-1}\gr_a\Phi:
L^2(\dR,\goth(\gl),\nu) \longrightarrow L^2(\dR,\goth(\gl),\nu)$.
Using the representations
\bed
\Phi\gr_\ga\Phi^{-1} = \Phi\gr^p_\ga\Phi^{-1} +
\Phi\gr^{ac}_\ga\Phi^{-1}
\eed
and
\bed
\Phi\gr^{ac}_\ga\Phi^{-1} = \Phi W_-\gr^{ac}_DW^*_-\Phi^{-1} = 
\Phi W_-\Psi^{-1}\Psi\gr^{ac}_D\Psi^{-1}\Psi W^*_-\Phi^{-1},
\eed
$\gr^{ac}_D := \gr_a \oplus \gr_b$, we get
\bed
M(\tilde{\rho}^{ac}_\ga) = \Phi W_-\Psi^{-1}M(\tilde{\rho}^{ac}_D)\Psi W^*_-\Phi^{-1}.
\eed
where
\be\label{apatra1}
\tilde{\rho}^{ac}_D(\gl) := 
\begin{cases}
0, & \gl \in \dR \setminus \gs_{ac}(H)\\
\gotf_b(\gl -\mu_b) & \gl \in [v_b,v_a)\\
\begin{pmatrix}
\gotf_b(\gl - \mu_b) & 0\\
0 & \gotf_a(\gl - \mu_a
\end{pmatrix}, & \gl \in [v_a,\infty).
\end{cases}
\ee
Taking into account \eqref{2.31} we prove \eqref{3.114}.
\end{proof}

\subsection{The case of sudden coupling}

Let us compare our results with following model \cite{AJPP}.
Assume that our system is not coupled for $t < 0$ and
suddenly at $t = 0$ the system becomes fully coupled. In a more
mathematical manner this can be modeled by the following family of
self-adjoint operators:
\bed
\widetilde{H}(t) :=
\begin{cases}
H_D & t < 0\\
H   & t \ge 0.
\end{cases}
\eed
To the evolution equation
\bed
i\frac{\partial}{\partial t}u(t) = \widetilde{H}(t)u(t),\quad t \in
\dR,
\eed
it corresponds a unique unitary solution operator or propagator
$\{\widetilde{U}(t,s)\}_{(t,s) \in \dR \times \dR}$ given by
\bed
\widetilde{U}(t,s) := \widetilde{U}(t)\widetilde{U}(s)^{-1}, \quad t,s \in \dR,
\eed
where
\bed
\widetilde{U}(t) = 
\begin{cases}
e^{-itH_D}, & t \le 0\\
e^{-itH},   & t > 0.
\end{cases}
\eed
The time evolution of the density operator is given by
\bed
{\gr}_\infty(t) :=
\widetilde{U}(t){\gr}_D\widetilde{U}(t)^*,
\quad t>0.
\eed
Clearly, $\lim_{t\to-\infty}\|{\gr}_\infty(t)
- \gr_D\| = 0$. Then using the identity:
\bed
\widetilde{\gr}(t) = e^{-itH}e^{itH_D}\gr_De^{-itH_D}e^{itH}, \quad t > 0,
\eed
we immediately get that
\bed
\slim_{t\to+\infty}{\gr}_\infty(t)P^{ac}(H) = W_-\gr_D W^*_-.
\eed
Hence we find
\bed
\slim_{T\to+\infty}\frac{1}{T}\int^T_0 dt\;{\gr}_\infty(t)P^{ac}(H) = W_-\gr_D W^*_-.
\eed
As above, we can show that
\bed
\slim_{T\to+\infty}\frac{1}{T}\int^T_0dt\;
P^{ac}(H){\gr}_\infty(t)P^s(H) = 0
\eed
and
\bed
\slim_{T\to+\infty}\frac{1}{T}\int^T_0dt\;
P^s(H){\gr}_\infty(t)P^s(H) = \sum^N_{j=1} E_H(\{\gl_j\})\gr_DE_H(\{\gl_j\}).
\eed
Hence we find
\bed
\slim_{T\to+\infty}\frac{1}{T}\int^T_0dt\;{\gr}_\infty(t) =
\sum^N_{j=1} E_H(\{\gl_j\})\gr_DE_H(\{\gl_j\}) + W_-\gr_DW^*_-.
\eed

\section{The stationary current, the Landau-Lifschitz and the 
Landauer-B\"uttiker formula}\label{section4}

There are by now several proofs of the Landauer-B\"uttiker formula
in the NESS approach (see \cite{AJPP,Ne1}), and in the finite volume
regularization approach (see \cite{CJM1,CJM2,CJM3}). Here we give
yet another proof in the NESS approach. In fact, we will only justify the
so-called Landau-Lifschitz current density formula (see
\eqref{atreia9} in what follows), which was the starting point in \cite{BKNR1}
for the proof of the Landauer-B\"uttiker formula (see Example 5.11 in that
paper).

Let us start by defining the stationary current, in the manner
introduced in \cite{AJPP}. Let $\eta >0$, and choose an 
integer $N\geq 2$. Denote by $\chi_b$ the characteristic function of the
interval $(b,\infty)$ (the right reservoir). Without loss of
generality, let us assume that $H>0$. 
\begin{defn}\label{defzece}
The trace class operator
\be\label{defopcur}
j(\eta):=i[H(1 +\eta H)^{-N},\chi_b]
\ee
is called the regularized current operator. The stationary current
coming out of the right reservoir is defined to be 
\be\label{defcur}
\mathfrak{I}_\alpha:=\lim_{\eta\searrow 0}\tr(\gr_\alpha j(\eta)).
\ee
\end{defn}
Now a few comments. The current operator is trace class because we can
write it as 
$$i[H(1 +\eta H)^{-N}-H_D(1 +\eta H_D)^{-N},\chi_b]$$
which clearly is trace class. Then since $\gr_\alpha^p$ does not
contribute to the trace in the definition of the current, we will
focus on the clearly $\alpha$ independent quantity:
\be\label{defcur2}
\mathfrak{I}=\lim_{\eta\searrow 0}\tr\left \{W_-\gr_DW_-^* P^{ac}(H)j(\eta)\right \}.
\ee

We start with a technical result:
\bl\label{lematrasa}
Let $\chi$ a bounded, compactly supported function. Then the operator 
$\chi (1+H)^{-2}$ is trace class.
\el
\begin{proof}
Choose a smooth and compactly supported function $\tilde{\chi}$ such
that $ \tilde{\chi}\chi=\chi$. Write 
$$\chi (1+H)^{-2}=\chi \tilde{\chi}(1+H)^{-2}=\chi
(1+H)^{-1}\tilde{\chi}(1+H)^{-1}+
\chi (1+H)^{-2}[H,\tilde{\chi}](1+H)^{-1}.$$
Since $\tilde{\chi}$ is smooth an compactly supported, the operator 
$(1+H)^{-1}[H,\tilde{\chi}](1+H)^{-1}$ is Hilbert-Schmidt. The
operators 
$\chi (1+H)^{-1}$ and $\tilde{\chi}(1+H)^{-1}$ are also
Hilbert-Schmidt, thus $\chi (1+H)^{-2}$ can be written as a sum of
products of Hilbert-Schmidt operators, therefore it is trace class.
\end{proof}

Next, let us now prove that we can replace the sharp characteristic
function in the definition of the current with a smooth one. Let $c>b+1$. 
Choose any function $\phi_c\in C^\infty(\mathbb{R})$
such that 
\be\label{fic}
0\leq \phi_c\leq 1,\quad \phi_c(x)=1\; {\rm if}\; x\geq c+1,\quad {\rm
  supp}(\phi_c)\subset (c-1,\infty).
\ee
Then let us prove the following identity: 
\bl\label{lemacons}
\bea\label{condervare}
\lefteqn{
\tr\left \{W_-\gr_DW_-^* P^{ac}(H)j(\eta)\right \}=}\\
& &
i\tr \left \{W_-\gr_DW_-^* P^{ac}(H)[H(1 +\eta H)^{-N},\phi_c]\right \}.
\nonumber
\eea
\el
\begin{proof}
First, the commutator $[H(1 +\eta H)^{-N},\phi_c]$ defines a trace
class operator; that is because now  
$[H,\phi_c]=-\frac{1}{2m_b}(\frac{d}{dx}\phi_c'+\phi_c'\frac{d}{dx})$,
and 
$(1 +\eta H)^{-1}[H,\phi_c](1 +\eta H)^{-1}$ is a trace class operator
(we can write it as a sum of products of two Hilbert-Schmidt
operators).  We also use the identity 
\be\label{intertz}
W_-\gr_DW_-^* P^{ac}(H)=W_-\gr_D(1+H_D)W_-^* P^{ac}(H)(1+
H)^{-1}
\ee
which is an easy consequence of the intertwining property of $W_-$.

Second, \eqref{condervare} would be implied by:
\be\label{condervare3}
\tr
\left \{W_-\gr_DW_-^* P^{ac}(H)[H(1 +\eta H)^{-N},\phi_c-\chi_b]\right \}=0.
\ee
We see that $\phi_c-\chi_b$ has compact support. If we write the
commutator as the difference of two terms, both of them will be trace
class. The first one is
\begin{align}
&W_-\gr_DW_-^* P^{ac}(H)H(1 +\eta H)^{-N}(\phi_c-\chi_b)\nonumber \\
&=
W_-\gr_D(1+H_D)^2W_-^* P^{ac}(H)H(1 +\eta
H)^{-N}\{(1+H)^{-2}(\phi_c-\chi_b)\}
\nonumber
\end{align}
and the second one is 
\begin{align}
&W_-\gr_DW_-^* P^{ac}(H)(\phi_c-\chi_b)H(1 +\eta
H)^{-N}\nonumber \\
&=W_-\gr_D(1+H_D)^2W_-^* P^{ac}(H)
\{(1+H)^{-2}(\phi_c-\chi_b)\}H(1 +\eta H)^{-N}.
\nonumber
\end{align}
Now according to Lemma \ref{lematrasa}, $(1+H)^{-2}(\phi_c-\chi_b)$ is a trace class
operator. Thus the two traces will be equal due to the
cyclicity property and the fact that $H$ commutes with the steady
state.
\end{proof}

We can now take the limit $\eta\searrow 0$:
\bl\label{lemaetazero}
The operator $(1+H)^{-2}[H,\phi_c]$ is trace class, and 
\begin{align}\label{atreia1}
\mathfrak{I}=i\tr\left \{W_-\gr_DW_-^* P^{ac}(H)[H,\phi_c]\right \},
\end{align}
independent of $\phi_c$.
\el
\begin{proof}
Note that $[H,\phi_c]=-\frac{1}{2m_b}(2\frac{d}{dx}\phi_c'-\phi_c'')$,
where both $\phi_c'$ and $\phi_c''$ are compactly
supported. Using the method of Lemma \ref{lematrasa} one can prove
that  $(1+H)^{-2}\frac{d}{dx}\phi_c'$ is trace class, hence
$(1+H)^{-2}[H,\phi_c]$ is trace class. Thus $W_-\gr_DW_-^*
P^{ac}(H)[H,\phi_c]$ is trace class since we can write 
$$W_-\gr_DW_-^* P^{ac}(H)[H,\phi_c]=W_-\gr_D(1+H_D)^2W_-^*
P^{ac}(H)(1+H)^{-2}[H,\phi_c].$$
In fact, using trace cyclicity one can prove that 
\bea\label{atreia2}
\lefteqn{
\tr\left \{W_-\gr_DW_-^* P^{ac}(H)[H,\phi_c]\right \}=}\\
& &
\tr\left \{W_-\gr_D(1+H_D)^2W_-^*
  P^{ac}(H)(1+H)^{-1}[H,\phi_c](1+H)^{-1}\right \}
\nonumber \\
& &
=-\tr\left \{W_-\gr_D(1+H_D)^2W_-^*
  P^{ac}(H)[(1+H)^{-1},\phi_c]\right \}.
\nonumber
\eea
This last identity indicates the strategy of the proof. Write:
\begin{align}\label{atreia3}
&\tr\left \{W_-\gr_DW_-^* P^{ac}(H)[H(1+\eta H)^{-N},\phi_c]\right
\}\\ &=
\tr\left \{W_-\gr_D(1+H_D)^3W_-^*
  P^{ac}(H)(1+H)^{-2}[H(1+\eta H)^{-N},\phi_c](1+H)^{-1}\right
\}\nonumber .
\end{align}
Now it is not so complicated to prove that 
$(1+H)^{-2}[H(1+\eta H)^{-N},\phi_c](1+H)^{-1}$ converges in the trace
norm to $(1+H)^{-2}[H,\phi_c](1+H)^{-1}$ when $\eta\searrow 0$; we do
not give details. Now use \eqref{condervare} and take the limit; we
obtain:
\begin{align}\label{atreia4}
\mathfrak{I}&=
i\tr\left \{W_-\gr_D(1+H_D)^3W_-^*
  P^{ac}(H)(1+H)^{-2}[H,\phi_c](1+H)^{-1}\right
\}\nonumber \\
&=i\tr\left \{W_-\gr_DW_-^*
  P^{ac}(H)[H,\phi_c]\right
\},
\end{align}
where in the last line we used trace cyclicity.
\end{proof}

The remaining thing is to compute the trace in \eqref{atreia1} using
the spectral representation of $H$. Let us compute the integral kernel
of $\mathcal{A}:=iW_-\gr_DW_-^*
  P^{ac}(H)\frac{1}{2m_b}\left
    (-\frac{d}{dx}\phi_c'-\phi_c'\frac{d}{dx}\right )$
in this representation.  We use \eqref{apatra1}, where 
we denote the diagonal elements of $\tilde{\gr}_D^{ac}(\lambda)$ by
$\tilde{\gr}_D^{ac}(\lambda)_{pp}$ (the other entries are zero). We obtain:
\bea\label{atreia7}
\lefteqn{
\mathcal{A}(\lambda,p;\lambda',p')=}\\
& &
-\frac{i}{2m_b}
\tilde{\gr}_D^{ac}(\lambda)_{pp}\int_{\mathbb{R}}\overline{\tilde{\phi}_p(x,\lambda)}
\left (\frac{d}{dx}\phi_c'(x)+\phi_c'(x)\frac{d}{dx}\right
)\tilde{\phi}_{p'}(x,\lambda')dx
\nonumber \\
& &
=-\frac{i}{2m_b}
\tilde{\gr}_D^{ac}(\lambda)_{pp}\int_{\mathbb{R}}\phi_c'(x)\{
\overline{\tilde{\phi}_p(x,\lambda)}\tilde{\phi}_{p'}'(x,\lambda')-
\overline{\tilde{\phi}_p'(x,\lambda)}\tilde{\phi}_{p'}(x,\lambda')\}dx,
\nonumber
\eea
where in the second line we integrated by parts (remember that
$\phi_c'$ is compactly supported). 

In order to compute the trace, we put $\lambda=\lambda'$, $p=p'$, and
integrate/sum over the variables. We obtain:
\begin{align}\label{atreia8}
\mathfrak{I}=\int_{\mathbb{R}}\phi_c'(x)j(x)dx,
\end{align}
where 
\begin{align}\label{atreia9}
j(x):=\frac{1}{m_b}\int_{v_b}^\infty \sum_p
\tilde{\gr}_D^{ac}(\lambda)_{pp}\Imag \{
\overline{\tilde{\phi}_p(x,\lambda)}\tilde{\phi}_{p}'(x,\lambda)\}d\lambda
\end{align}
is the current density, which can be shown to be independent of $x$
(the above imaginary part is a Wronskian of two solutions of a Schr\"odinger equation, see
\cite{BKNR1} for details). But $\int_{\mathbb{R}}\phi_c'(x)dx=1$ for our class of
cut-off functions, therefore the stationary current equals the
(constant) value of its density. The Landauer-B\"uttiker formula
follows from the Landau-Lifschitz formula \eqref{atreia9} as proved in \cite{BKNR1}. 

\begin{acknowledgements}
H.C. and H.N. acknowledge support from the Danish 
F.N.U. grant {\it  Mathematical Physics and Partial Differential
  Equations}. This work was initiated during a visit of H.C. and V.Z. at WIAS, and they are 
thankful for the hospitality and financial support extended to them during the work on this paper.

\end{acknowledgements}

\end{document}